\documentstyle[psfig]{mn}

\newif\ifAMStwofonts
\newcommand{\be}{\begin{equation}}
\newcommand{\ee}{\end{equation}}
\newcommand{\bea}{\begin{eqnarray}}
\newcommand{\eea}{\end{eqnarray}}

\include{mn_h}

\title{Joint Cosmological
Formation of QSOs and Bulge-dominated Galaxies}

\author[P. Monaco, P. Salucci \& L. Danese]
  {Pierluigi Monaco,$^{1,2,3}$
   Paolo Salucci$^3$ and 
   Luigi Danese$^3$ \\
   $^1$Institute of Astronomy, Madingley Road, Cambridge CB3 0HA, GB\\
   $^2$Dipartimento di Astronomia, via Tiepolo 11, 34131 Trieste -- Italy\\
   $^3$SISSA, via Beirut 4, 34013 Trieste -- Italy}

 \date{}

\pagerange{\pageref{firstpage}--\pageref{lastpage}}
\pubyear{1999}

\begin{document}

\maketitle

\label{firstpage}

\begin{abstract}
Older and more recent pieces of observational evidence suggest a
strong connection between QSOs and galaxies; in particular, the
recently discovered correlation between black hole and galactic bulge
masses suggests that QSO activity is directly connected to the
formation of galactic bulges.  The cosmological problem of QSO
formation is analyzed in the framework of an analytical model for
galaxy formation; for the first time a joint comparison with galaxy
and QSO observables is performed.  In this model it is assumed that
the same physical variable which determines galaxy morphology is able
to modulate the mass of the black hole responsible for QSO activity.
Both halo spin and the occurence of a major merger are considered as
candidates to this role.  The predictions of the model are compared to
available data for the type-dependent galaxy mass functions, the
star-formation history of elliptical galaxies, the QSO luminosity
function and its evolution (including the obscured objects
contributing to the hard-X-ray background), the mass function of
dormant black holes and the distribution of black-hole -- bulge mass
ratios.  A good agreement with observations is obtained if the halo
spin modulates the efficiency of black-hole formation, and if the
galactic halos at $z=0$ have shone in an inverted order with respect
to the hierarchical one (i.e., stars and black holes in bigger
galactic halos have formed before those in smaller ones).  This
inversion of hierarchical order for galaxy formation, which reconciles
galaxy formation with QSO evolution, is consistent with many pieces of
observational evidence.

\end{abstract}


%%%%%%%%%%%%%%%%%%%%%%%%%%%%%  1  %%%%%%%%%%%%%%%%%%%%%%%%%%%%%%%%%%
\section{Introduction}

High-redshift quasi-stellar objects (QSOs) have been for a long time
the only probe to the high-redshift Universe at $1\la z\la 4$.
However, their potential power in constraining cosmological models has
always been hampered by their complexity as a physical phenomenon:
they are thought to be powered by huge black holes (BHs), of mass
$\sim 10^{6} - 10^{10} M_\odot$, hosted at the center of
proto-galactic halos, so that their activity couples very different
scales, from fractions of pc to Mpc.  The problem is then unsuitable
for numerical studies, and the modeling of QSOs in a cosmological
framework must rely on analytical approximations, typically based on
uncertain hypotheses or poorly constrained parameters.

The QSO population is thought to be made of many generations of
short-lived events, with lifetimes ranging from a few $10^7$ to $10^8$
yr, as longer lifetimes would imply few very large dormant BHs, with
masses $\sim 10^{12} M_\odot$, which is contrary to the observational
evidence (see, e.g., Cavaliere \& Padovani 1988).  This fact has two
implications, which reveal the deep connection between QSOs and
galaxies: firstly, the number density of expected dormant objects
matches that of bright galaxies; secondly, to evolve on cosmological
time scales as observed, the various generations of QSOs must be
coordinated by a process ({\it great coordinator}), which is likely to
be that of galaxy formation.

In fact, while QSOs have always been addressed as a separate field in
cosmology, both theoretically and observationally, their connection
with galaxies and galaxy formation is receiving ever growing evidence
from low and high redshift observations.  There is no evidence of QSO
activity outside galaxies, and the direct observation of high-redshift
galaxies has revealed that large spheroidal galaxies are the most
common hosts for bright QSOs (see, e.g., Hall \& Green 1998; McLure et
al. 1998).  Massive dark objects, interpretable as large dormant BHs,
are routinely found in nearby spheroids (Kormendy \& Richstone 1995;
Ford et al. 1997; Magorrian et al. 1998; van der Marel 1998; Ho 1998;
Wandel 1998).  The estimates of their masses are still affected by
many systematics, but two resulting evidences seem robust: (i) the
mass of the massive dark object is correlated to the mass of the bulge
component, (ii) this correlation has a scatter of about one order of
magnitude.  Finally, the light history of QSOs has an interesting
resemblance with the star-formation history of galaxies (Boyle \&
Terlevich 1998; Cavaliere \& Vittorini 1998).

The presence of these large dormant BHs in the cores of the bulges of
nearby galaxies is a key quantity for testing the BH paradigm of QSOs
(Soltan 1982).  A previous paper (Salucci et al. 1998a; hereafter
paper I) has been dedicated to finding the mass function of dormant
BHs, using up-to-date observations and including the contribution of a
population of heavily obscured objects (type II AGN) which are
revealed by their contribution to the hard-X-ray background (see also
Iwasawa \& Fabian 1998).  The mass function of dormant BHs in nearby
galaxies has been estimated using two methods. The first exploits the
recently discovered correlation of BH and bulge masses, and consists
in convolving the mass function of the galactic bulges with a BH-bulge
relation inferred from observation. The second method relies on the
correlation between radio power and BH mass (see also Franceschini,
Vercellone \& Fabian 1998). These two estimates agree with the mass
function of the accreted matter inferred from the QSO emission
(including obscured objects), giving direct support to the QSO galaxy
connection. The local mass density in BHs turnes out to be $\sim 6.5\
10^5\ M_\odot$ Mpc$^{-3}$ for $H_0=70$ km/s/Mpc

The results found in paper I reveal a dichotomy between large BHs
($M>10^8\ M_\odot$) and small BHs.  The former are hosted in
elliptical galaxies, shine only once near to the Eddington rate, and
tend to be not obscured, while the latter are found in the bulges of
spiral galaxies, shine with a lower efficiency, can be reactivated by
interactions and are frequently obscured.  In Salucci et al. (1998b)
the mass function of BHs in spirals has been constrained through the
use of several hundred rotation curves of spirals; no BH is detected,
and the upper limits thus obtained constrain the numerous late spirals
to host a negligible amount of mass in BHs.

Within the framework of hierarchical cosmological models, it is
possible to make prediction on the number and properties of
dark-matter (DM) halos. Besides, to predict the statistical properties
of QSOs it is necessary to make assumptions on the probability that a
BH forms inside a DM halo, and the efficiency with which such a BH
radiates energy.  Arguments in favor of BH formation in normal
galactic halos are given, for instance, in Rees (1984) and Haehnelt \&
Rees (1993).  Efstathiou \& Rees (1988) made the assumption of
constant ratio between halo and BH mass and of radiation at the
Eddington limit, to conclude that the standard cold dark matter (CDM)
model could reproduce the QSO luminosity function.  A similar but more
refined procedure was used by Haehnelt \& Rees (1993): they assumed
that BHs form with an efficiency which increases with central halo
density and halo virial velocity, so as the QSO activity reaches a
maximum at $z\sim 3$ as observed.  Haehnelt, Natarajan \& Rees (1997)
considered the case in which most mass is accreted during the
quiescent phase, while Cattaneo, Haehnelt \& Rees (1999) analyzed the
implications of the BH-bulge correlation, especially on galaxy
mergings.  A related approach was used by Katz et al. (1994), who used
N-body simulations to check whether the CDM model could give a
sufficient number of suitable halos to justify QSOs at $z\ga 3$.
Carlberg (1990), and more recently Krivitsky \& Kontorovich (1998),
estimated the number of QSOs by assuming them to be related to galaxy
mergings.  Predictions on the cosmological evolution of QSOs were
given by Haiman \& Menou (1998) and Percival \& Miller (1999).
Eisenstein \& Loeb (1995a,b) followed in some detail the dissipation
of angular momentum of gas infalling inside a DM halo, modeling the
collapsing halos as homogeneous ellipsoids and seeking for suitable
conditions for a BH to form: they concluded that seed BHs (with mass
$\sim 10^5\ M_\odot$) can form at rather large redshifts ($z>5$),
giving rise to QSO activity when included in a protogalaxy (see also
Loeb 1993).  Cavaliere \& Vittorini (1998) argued that QSOs at $z\ga
3$ are connected to the formation of new galactic halos, while the
already-formed BHs are reactivated by galaxy interactions in
newly-formed groups, a mechanism which is effective at $z\la 3$ (see,
e.g., Monaco et al. 1994, and references therein).

This paper aims to construct, for the first time, an analytical model
for joint QSO and galaxy formation, which reproduces the main
observables relative to both populations.  Mergers and/or halo spin
are proposed as possible physical variables responsible for both
galactic morphology and BH formation.  The analytical model is
designed to reproduce the mass function of galactic halos for
different broad morphological classes, the star formation history of
ellipticals, the QSO luminosity function and its redshift evolution,
the mass function of dormant BHs and the BH-bulge relation.  The plan
of the paper is as follows: Section 2 presents the analytical model
for galaxy formation, based on the joint distribution of halo mass,
spin or last merger, and formation time.  In Section 3 the model is
shown to reproduce the estimated halo mass function of galaxies,
divided into broad morphological classes, and the star-formation
history of ellipticals.  In Section 4 the model is compared to the
observed QSO luminosity function.  Section 5 addresses the dormant BH
masses and their relation to bulge masses.  Section 6 contains a
summary and some final remarks.

A Hubble constant of $H_0 = 50\ km\ s^{-1}\ Mpc^{-1}$ will be used
when discussing Einstein-de Sitter (EdS) models, while a value of 70
will be used for the open model or the model with cosmological
constant.

%%%%%%%%%%%%%%%%%%%%%%%%%%%%%  2  %%%%%%%%%%%%%%%%%%%%%%%%%%%%%%%%%%
\section{An analytical model for galaxy and QSO formation}

Galaxy formation in hierarchical models is usually addressed by means
of semi-analytical techniques, in which the abundance of DM halos and
their merging histories are inferred either from N-body simulations or
from the extended PS formalism (Bond et al. 1991; Bower 1991; Lacey \&
Cole 1993), and the history of gas in halos is described through a set
of simplified rules for gas cooling, star formation, feedback
processes, galaxy mergings etc. (see, e.g, White 1993; Cole et
al. 1994).  Noteworthy, the main qualitative conclusions of these
semi-analytic models can be reached by means of simple analytic
arguments based on the PS mass function (White \& Rees 1978; White \&
Frenk 1991).  It is then reasonable, at this stage, to construct a
simple analytical model for joint galaxy and QSO formation, leaving
detailed calculations to further analysis.

The power spectra considered in the present paper are the CDM-like
ones described by the parameterization of Efstathiou, Bond \& White
(1993), with the shape parameter defined as $\Gamma=\Omega h$, where
$\Omega$ is the cosmological density parameter and $h=H_0/(100\
km/s/Mpc)$.  Three models have been analyzed, namely an EdS model with
$\Omega=1$ and $h=0.5$ (so that $\Gamma=0.5$), a flat low-density
model with $\Omega=0.3$ and cosmological constant ($h=0.7$ and
$\Gamma=0.21$), and an open model with $\Omega=0.3$ and $h=0.7$
($\Gamma=0.21$).  Following Eke et al. (1998), the normalization has
been fixed so that the standard deviation of the initial density
(linearly extrapolated to $z=0$) on a 8 $h^{-1}$ Mpc sphere,
$\sigma_8$, is 0.7 for the EdS model, and 1 in the other cases.  For
the sake of brevity, the results of the open model, which are similar
to the cosmological constant case (hereafter called Lambda model), are
not shown.

\subsection{Dark-matter halos in hierarchical Universes}

The mass function of DM halos at a given redshift is reproduced by the
well-known PS formula (see Monaco 1998 for a recent review on the mass
function):

\bea \lefteqn{n_{\rm PS} (M_H;z)dM_H = }\label{eq:ps} \\ &&
\frac{\rho_0}{M_H}\left[\frac{1}{\sqrt{2\pi\Lambda^3}}
\exp\left(-\frac{\delta_c^2/b(z)^2}{2\Lambda}
\right)\right]\left|\frac{d\Lambda}{dM_H} \right| dM_H, \nonumber
\eea

\noindent
where $M_H$ is the halo mass, $\rho_0$ is the background density,
$\Lambda(M_H) \equiv \langle\delta^2\rangle$ is the mass variance at
the scale corresponding to $M_H$ (it is usually denoted as
$\sigma^2$), and $\delta_c$ is a threshold parameter.  Following
Monaco (1998, 1999) and Governato et al (1998), the $\delta_c$
parameter is set to 1.5 if the cosmological density is $\Omega=1$, and
1.69 if $\Omega<1$ (with or without cosmological constant).  The
time-dependent function $b(z)$ is the linear growing mode, normalized
as $b(z=0)=1$.  This quantity is related to the scale factor
$a(z)=1/(1+z)$ as follows: $b(z)\propto a(z)$ at high redshift, and
$b(z)=a(z)$ in the $\Omega=1$ cosmology.  The expressions for $b(z)$
for cosmologies different from EdS are given, for instance, in
Monaco (1998).  The critical mass $M_*(z)$ at a given redshift is
defined as the mass at which the variance (linearly extrapolated to
$z$) is equal to $\Lambda(M_*(z))=(\delta_c/b(z))^2$.

Galaxies do not form in every DM halo: large halos have large cooling
times, which can become larger than the dynamical time, in which case
baryons cannot gather into a single unit.  Following White \& Rees
(1978) and White (1993), the virial temperature $T_H$ of a halo of
mass $M_H$ and density $\varrho_H$ is assumed to scale as $T_H \propto
M_H^{2/3} \varrho_H^{1/3}$, while the cooling time depends on the
efficiency of cooling $\Lambda_{\rm cooling}(T_H)$ in the following
way: $t_{\rm cool} \propto T_H \varrho_H^{-1} \Lambda_{\rm
cooling}(T_H)^{-1}$.  In the relevant range of temperature, from $\sim
10^4$ to $\sim 10^{5.5} K$, the cooling efficiency scales roughly as
$\Lambda_{\rm cooling}(T_H) \propto T_H^{-1/2}$; then the cooling time
scales as $t_{\rm cool} \propto M_H \varrho_H^{-1/2}$.  Finally, the
dynamical time of the halo scales as $t_{\rm dyn}\propto
\varrho_H^{-1/2}$.  Then, a condition $t_{\rm cool}<t_{\rm dyn}$
translates into a redshift-independent cutoff of the mass function.

The largest halo mass in which gas can effectively cool down is called
$M_{\rm cool}$, and left as a free parameter.  Assuming that galaxies
obey a relation $M_H \propto L^{\beta}$, with $\beta\sim0.5-1$ (see
Section 3.1; $\beta$ depends on the morphological type, we use the
value relative to elliptical galaxies), we model the astrophysical
cutoff so as to reproduce the observed cutoff in the galaxy luminosity
function:

\be C_{\rm cool}(M_H) = \exp(-(M_H/M_{\rm cool})^{1/\beta}).
\label{eq:cutoff} \ee

This astrophysical cutoff is not the only feature that distinguishes
the assembly of galaxies from that of DM: the merging of baryonic
matter is regulated both by cooling and feedback, when gaseous clumps
merge, and by non-dissipative merging of clumps made of stars.  To
address the formation of a galaxy, it is necessary to distinguish
between early-forming halos, whose density is high enough for
subclumps to merge into a single galaxy, and late-forming halos, which
host small groups of galaxies (see Section 2.4).  We consider also a
further parameter $p$, responsible for morphology.  The number density
$n_H$ of halos with mass $M_H$, morphological parameter $p$ and
formation redshift $z_f$ can be expressed with great generality as
follows:

\bea \lefteqn{n_H(M_H,p,z_f) dM_H dp\, dz_f = n_{\rm PS}(M_H|z_f) \times
C_{\rm cool}(M_H) dM_H \times } \nonumber \\&& P_p(p|M_H,z_f) dp \times
P_f(z_f|M_H) dz_f.  \label{eq:joint_all} \eea

The following subsections are dedicated to finding suitable
approximations for the distributions $P_p(p|M_H,z_f)$, with $p$ equal
to the halo spin $\lambda$ or merging fraction $f$, and
$P_f(z_f|M_H)$.

\subsection{The joint mass-spin function}

Spin is acquired by a halo during the mildly non linear regime, when
the tidal coupling between the (proto-)halo and the external mass
distribution is effective. After decoupling, the halo inertia moment
becomes negligible, and spin stops growing, at least as long as the
structure remains isolated. It is then possible to give analytical
estimates of halo spins (Peebles 1969, White 1984, Heavens \& Peacock
1988, Eisenstein \& Loeb 1995a, Catelan \& Theuns 1996).  The spin
distribution of halos is also determined by means of N-body
simulations (Barnes \& Efstathiou 1988; Zurek, Quinn \& Salmon 1988;
Warren et al. 1992; Ueda et al. 1994; Lemson \& Kauffman 1998).  The
comparison of analytical and numerical estimates leads to the
following conclusions: (i) the spin distributions given by different
authors and with different methods are approximately consistent with
each other; (ii) the functional dependences of spin on other halo
parameters, predicted by means of analytical arguments, are consistent
with N-body results; (iii) the spin distribution is wide, and nearly
lognormal in shape.

The dimensionless spin parameter $\lambda$ is defined as:

\be \lambda = L E^{1/2} G^{-1} M_H^{-5/2}, \label{eq:spin} \ee

\noindent 
where $L$ is the final angular momentum of the halo, $E$ is its total
energy, $G$ is the gravitational constant and $M_H$ is the halo mass.
The spin parameter $\lambda$ is nearly independent of everything,
except a weak dependence of its mean value $\tilde{l}\equiv
\langle\log\lambda\rangle$ on halo mass:

\be \tilde{l}(M_H/M_*(z)) = \tilde{l}_0 -\alpha_\lambda \log (M_H/M_*(z)).
\label{eq:spinmass}\ee

\noindent 
With $\tilde{l}_0 =\log(0.04)$, the average value of the spin
parameter, for a set of halos with mass not much smaller than $M_*$,
is about 0.05, the value given by N-body simulations.  The exponent
$\alpha_\lambda$ is in the range 0.1 -- 0.2; the value 0.15 will be
used in the following.  This trend is theoretically explained by a
dependence of angular momentum on the initial height of the peak from
which the structure comes from (Catelan \& Theuns 1996), and has been
revealed in the N-body simulations of Ueda et al. (1994) and Cole \&
Lacey (1996).  The minus sign implies that rare, massive halos
(relative to $M_*(z)$) tend to have a lower spin.

The PDF of the $\lambda$ parameter, as given by Barnes \& Efstathiou
(1988), Warren et al. (1992), Ueda et al. (1994), Eisenstein \& Loeb
(1995), Catelan \& Theuns (1996) and Lemson \& Kauffman (1997), is
well approximated by the following lognormal distribution:

\be P_\lambda(\lambda|M_H,z) d\lambda = \frac{1}{\sqrt{2\pi\sigma_\lambda^2}}
\exp\left(-\frac{(\log\lambda-\tilde{l})^2} {2\sigma_\lambda^2}\right)
d\log \lambda, \label{eq:spinpdf} \ee

\noindent 
where $\sigma_\lambda=0.3$ and $\tilde{l}$ is given by
Eq.~\ref{eq:spinmass}.  Note that some authors, as Mo, Mao \& White
(1998), use for $\sigma_\lambda$ a value of 0.21, which is 30\%
smaller than the best-fit one used here.

\subsection{The joint mass-merging function}

Merging histories of DM halos are correctly reproduced by means of an
extension of the PS approach to the mass function problem (Bond et
al. 1991; Bower 1991; Lacey \& Cole 1993).  The probability that a
halo of mass $M_1$ (corresponding to a variance $\Lambda_1$) is
included into a halo of mass $M_2$ (corresponding to a variance
$\Lambda_2$) after a time interval $\Delta \log b$, where the growing
mode $b(z)$ is used as time variable, is:

\bea \lefteqn{P(M_1\rightarrow M_2;z)dM_1 \Delta\log b = 
\frac{1}{\sqrt{2\pi}} 
\frac{\delta_c}{b(z)} (\Lambda_1-\Lambda_2)^{-3/2}\times}  \nonumber 
\\&& \exp\left[
-\left(\frac{\delta_c}{b(z)}\right)^2 \frac{(\Delta\log b)^2}
{2(\Lambda_1-\Lambda_2)}\right] d\Lambda_1 \Delta\log b.
\label{eq:merging}\eea

\noindent 
Denoting the merging fraction $f$ as the ratio between the masses of
the merging and the final clumps, $f=M_1/M_2$, it is straightforward
from Eq.~\ref{eq:merging} to obtain the probability $P_f(f;M_H,z)$
that a halo (with final mass $M_H=M_2$) at redshift $z$ has
experienced a merging, within a time interval $\Delta\log b$, with
another halo of mass $(M_2-M_1)$.  In the following, $M_1$ will denote
the major progenitor of the halo, so that $M_1 \ge M_2/2$ and $1/2\ge
f \ge 1$.  The timescale $\Delta\log b$ is left as a free parameter.

\begin{figure}
\centerline{
\psfig{figure=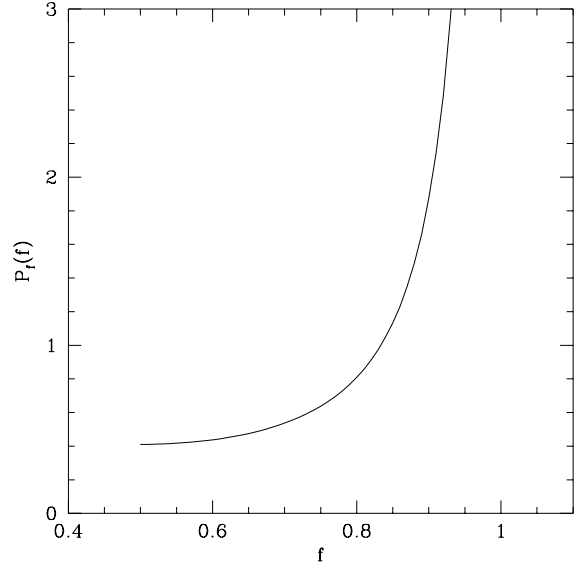,width=8cm}
}
\caption{The $P_f(f)$ distribution derived from Eq.~\ref{eq:merging},
for $M_H=M_*$ and $\Delta\log b = 0.1$.  The merging fraction $f$ is
the ratio between the masses of the two merging clumps (larger over
smaller).}
\end{figure}

The function $P_f(f;M_H,z)$ is straightforward to obtain, but its
expression is cumbersome; it is shown in Fig. 1 in the case $M_H=M_*$
and $\Delta\log b = 0.1$.  For $f\rightarrow 0.5$ the probability does
not vanish but saturates to a finite value, with a very flat shape,
which is not unexpected as the merging of two comparable masses is not
a rare event.  In Section 2.6 it will be shown that this is a problem
for using the merging fraction as a variable which modulates the
efficiency of BH formation.

In the simple case of power-law power spectra, $P(k)\propto k^n$, the
probability of having had a major merging, $f<f_c$ (where $f_c$ is a
threshold) depends on $M_H/M_*(z)$ in the following way:

\bea \lefteqn{P(f<f_c;M_H,z)\propto \left(\frac{M_H}{M_*(z)}\right)^{(n+3)/6}
\times}\label{eq:major} \\&&
\exp\left[-{\rm const}\times \left(\frac{M_H}{M_*(z)}\right)^{(n+3)/3}
\times \Delta\log b\right]. \nonumber\eea

\noindent
Eq.~\ref{eq:major} shows that the probability of having experienced a
mayor merging event grows with mass, in a way similar to that of the
spin parameter: the exponent is small again, of order 0.3 if the
spectral index $n$ is about $-2$.  At variance with the spin case,
this trend is not univocal: very large halos, with masses $M_H\gg M_*$
do not experience major mergings (they are just rare!); this fact does
not have a great importance, as these very rare objects turn out to be
mostly irrelevant.

\subsection{The redshift of halo formation and shining}

The assembly of baryonic structures like galaxies does not follow the
assembly of DM halos: a DM halo can host more than one galaxy, and
galaxies inside DM halos can subsequently merge.  At the present time,
most galaxies are not isolated structures, but are contained in larger
halos, such as groups and clusters.  As a consequence, most
present-day galactic halos are not related to the isolated halos
described by the PS mass function; those halos have formed, and were
present ``in the PS sense'' at high redshift, and have subsequently
gathered into larger structures.  The cores of such halos have
retained their identity as they had much higher densities than the
groups in which they have fallen.

In order to describe the formation of such galactic halos, it is
necessary to have some information on their dynamical history.  This
information is not contained in the simple PS mass function, but can
be obtained in the extended PS formalism, for instance with the
semi-analytical merging trees technique (Lacey \& Cole 1993), in which
the merging histories of a sample of halos are given.  A simpler,
analytical procedure was recently proposed by Percival \& Miller
(1999), who tested its validity against large N-body simulations.  The
extended PS theory gives the probability $P(M_H|z)$ that a halo of
mass $M_H$ is present at a given redshift $z$; it is possible to
invert such probability to obtain the probability $P_f(z_f|M_H)$ that a
halo of mass $M_H$ forms at the redshift $z_f$:

\be P_f(z_f|M_H) dz_f = \frac{\delta_c^2}{\Lambda b^2}
\exp\left(-\frac{\delta_c^2}{2\Lambda b^2}\right)
\frac{1}{b}\frac{db}{dz} dz_f.
\label{eq:perc} \ee

As in Section 2.3, the variance $\Lambda(M_H/M_*(z))$ yields a
dependence on the halo mass in units of the critical mass $M_*(z)$.
As the curve is peaked on $\Lambda\sim 1$, objects form preferentially
when their mass is not very different from $M_*$, and then larger
masses form at larger times.  Another characteristic of
Eq.~\ref{eq:perc} is that a large number of halos of mass $\sim
10^{12}M_\odot$ are predicted to form at small redshift, $z_f<1$.
Such objects can be either galaxies or small groups, depending on
whether the baryonic substructures manage to merge into a single
entity.  The cross section for dissipationless merging is proportional
to the square of the halo density, which scales as the cosmological
density.  As a consequence, the probability that a halo is going to
host a single galaxy (at $z$=0) is suppressed at lower redshift as
$(1+z_f)^6$.  Then, in order to pick up the galactic halos, the
Percival \& Miller (1999) probability of formation redshift is
multiplied by:

\be C_{\rm dens} = \left(1+\left(\frac{1+z_0}{1+z_f}\right)^6\right)^{-1}.
\label{eq:dens} \ee

\noindent
The reference redshift $z_0$ is left as a free parameter.  It is
useful to stress that this condition, related to dissipationless
merging, is a complement to the cooling condition given in
Eq~\ref{eq:cutoff}.

\begin{figure*}
\centerline{
\psfig{figure=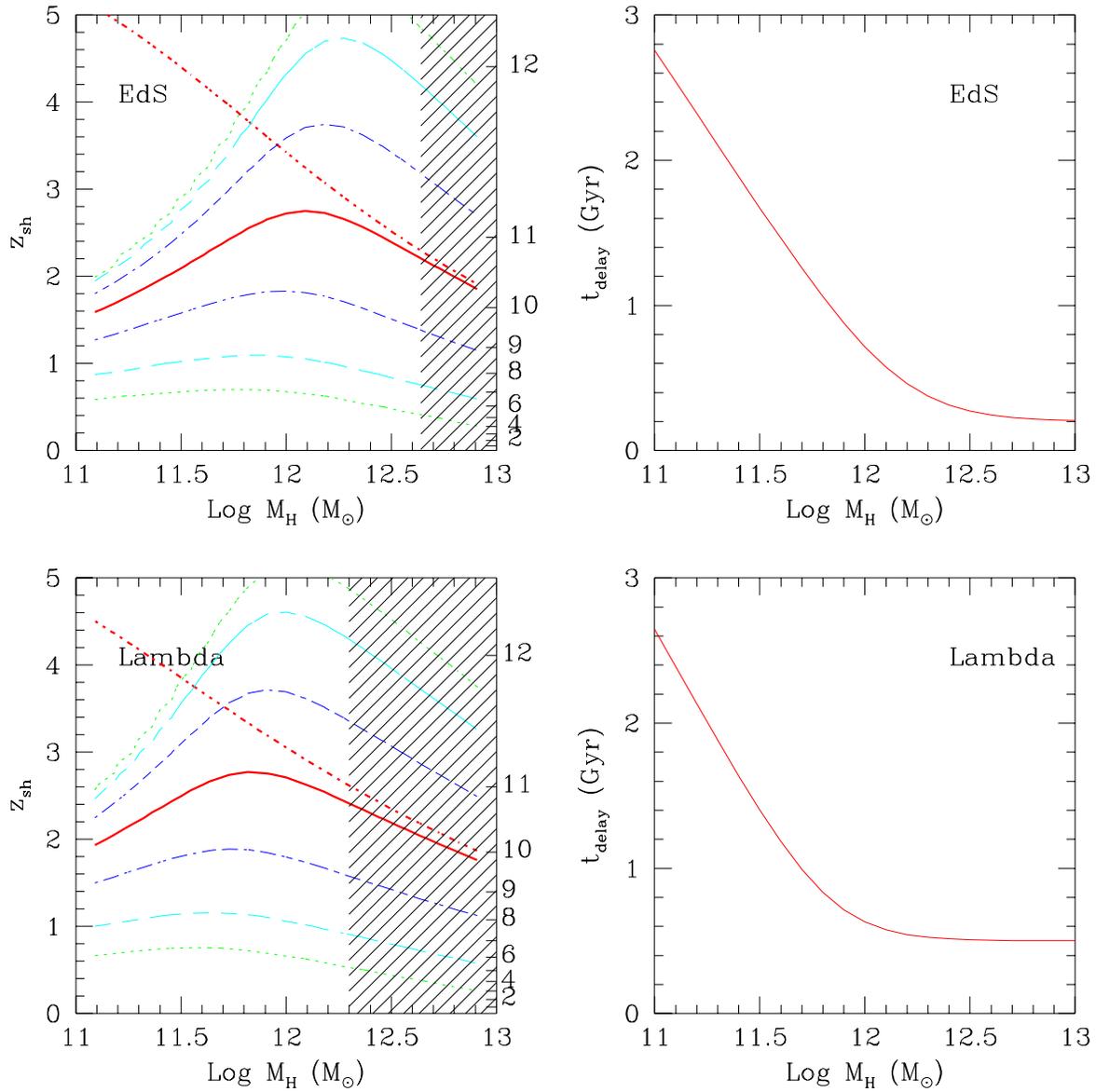,width=17cm}
}
\caption{Left panels: redshift of halo shining $z_{\rm sh}$ for
elliptical galaxies, as a function of the halo mass.  The lookback
time in Gyr is given on the right axis.  The thick continuous line
gives the redshift at which 50\% of halos have shone; the thick dashed
lines are what is obtained if there is no delay.  The dot-dashed,
dashed and dotted lines give the redshift interval within which 68\%,
95\% and 99\% of halos shine.  The shaded areas highlight those
galactic halos whose abundance is very small (their mass is larger
than 3 times the halo mass corresponding to an $L_*$ elliptical).
Right panels: delay time $t_{\rm delay}$ as a function of halo mass
$M_H$.  The Eds (upper panels) and Lambda (lower panels) models are
shown.}
\end{figure*}

The formation of stars in spheroids and the bright phase of QSOs are
assumed to be close in time.  The high metallicities inferred from QSO
spectra suggest that significant star formation in the host galaxies
has already taken place before the bright QSO phase (Hamann \& Ferland
1993).  Besides, QSO activity cannot easily have place much later than
the formation of bulge stars, when cold gas is almost absent and the
feeding of BHs is difficult.

QSO activity does not take place immediately at the dynamical
formation time of a generic (galactic) DM halo.  We assume that the
`shining phase'\footnote{In the present paper the phrases `shining of
the halo' and `galaxy formation' are used as synonyms.} of the halo,
i.e. when QSO activity takes place and star formation is already
turned on, is delayed by a time $t_{\rm delay}$ with respect to the
dynamical formation of the halo.  This delay is assumed to be longer
for smaller halos, such as to invert the hierarchical order for halo
shining.  In this case, the brighter QSOs can shine before the fainter
ones, as observed.  An alternative scenario able to reproduce the QSO
luminosity evolution requires lower efficiency of BH formation with
increasing time and, as a consequence, in larger halos (Haehnelt \&
Rees 1993).  This would induce an anti-correlation between bulge and
BH mass, contrary to the observational evidence.

As mentioned above, metallicity studies of the QSO environments show
that significant star formation in the host galaxies has occurred
before the QSO shining phase.  Thus we can infer that star formation
in larger early-type galaxies is turned on more rapidly.  In other
words, the QSO evolution marks the history of the star formation rate
in early type galaxies (Silva et al. 1999).  Shorter time scales for
star formation in massive ellipticals have been suggested by Matteucci
(1994) and Bressan, Chiosi \& Tantalo (1996), on the basis of the
chemical evolution of star populations.

The shorter delay of the QSO shining phase for larger BHs may be
ascribed to several mechanisms.  For instance, more powerful objects
can be able to remove earlier, and from a larger solid angle, the dust
surrounding the circumnuclear regions.  The gas in larger galaxies may
gather more easily in the core of the host halo, because of lower
angular momentum, and cool down, thus accreting on a seed BH.  Smaller
halos may accrete gas later in secondary infall events.  An additional
possibility is that large halos are endowed of dense peaks which may
merge rapidly to produce supermassive BHs.

Since the details of processes involving baryons are not fully
understood, we presently explore the mass-dependent delay hypothesis
as a heuristic (and parametric) guess, which helps to reconcile the
`anti-hierarchical' evolution of QSOs with the bulge-BH relation, and
is consistent with evidences coming from the study of stellar
populations in ellipticals.

If $t_f=t(z_f)$ is the dynamical formation time, the ``shining'' time
of the halo is $t_{\rm sh}=t_f+t_{\rm delay}$, and its ``shining''
redshift $z_{\rm sh}=z(t_{\rm sh})$.  The halo mass $M_H$ will denote
in the following the mass of the halo at the shining time $t_{\rm
sh}$; the number density of halos will be calculated at the shining
redshift.  We have found it convenient to parameterize the dependence
of the delay time on the halo mass as follows:

\be t_{\rm delay}(M_H) = \log\left(10^{t_f-\alpha_f(\log M_H - \log
M_{H*}^E)} + 10^{t_f}\right).
\label{eq:delay} \ee

\noindent
Here $t_f$ is the delay of a galaxy of halo mass larger than
$M_{H*}^E$, corresponding to an $L_*$ elliptical (see Section 3.1).
The parameterization is such that while large halos are delayed by
$t_f$, smaller ones are delayed proportionally to $\log
(M_H/M_{H*}^E)$.  Eq.~\ref{eq:perc}, multiplied by the cutoff given by
Eq.~\ref{eq:dens}, is evaluated at the shining redshift $z_{\rm sh}
=z(t_{\rm sh})$; in other words, the curve is shifted in time by an
amount $t_{\rm delay}$.

Notably, the final number of objects is calculated by integrating the
contributions at various redshifts.  This is not strictly correct, as
small halos at a time can be part of larger halos at a following time;
this introduces an uncertainty in the normalization of the number of
halos.  However, the effect is likely to be modest, especially if
large halos shine before smaller ones.  The use of semi-analytical
techniques, based on the merging trees, would solve this problem, and
would allow a more detailed description of the delay time.  Moreover,
it would allow to relax the assumption, implicit in this approach,
that the merging of the halos of already-formed spheroids is
negligible.  As mentioned at the beginning of this Section, this
analytical approach is supposed to catch the most important
dependences, while further refinements are left to future work.

To quantify and visualize the inversion of hierarchical order, it is
useful to consider how the abundance of galactic halos of fixed mass
grows in time.  This is not the same as plotting Eq.~\ref{eq:perc},
which does not take into account the different abundance of halos of
given mass at different times.  The abundance of galactic halos is
presented below (Eq.~\ref{eq:galaxies}).  The left panels in Fig. 2
show the distribution of shining redshifts $z_{\rm sh}$ for halos of
fixed mass which host ellipticals (with spin threshold) at $z=0$, for
the EdS and Lambda cosmologies.  The thick line shows the redshift at
which half of the $z=0$ halos have shined, the other lines show the
redshift intervals within which 68\%, 95\% and 99\% of halos shine.
The right panels show the delay time $t_{\rm delay}(M_H)$ used.  The
best-fit parameters have been used, see Table 2 and Section 2.7.  For
comparison, the thick dashed line shows the 50\% line obtained by
assuming no delay.  The inversion of hierarchical order is visible in
the change of slope of the ``iso-shining'' lines at moderate and small
halo masses.

Fig. 2. predicts that many big ellipticals form at high redshift;
because of bias, these will preferentially end up in clusters.  This
is consistent with the observational evidence, based on the
colour-magnitude relation (Bower, Lucey \& Ellis 1992; Ellis et
al. 1997; Kodama et al. 1998; but see also Shioya \& Bekki 1998), the
tightness of the fundamental plane (Renzini \& Ciotti 1993; van Dokkum
et al. 1998), and the $Mg_b- \sigma_0$ relation (Ziegler \& Bender
1997), that cluster ellipticals form a homogeneous class of old
objects.  Pushing the observations to high redshift clusters allows to
tighten the constraint, but only for big objects, not much smaller
than $10^{12}\ M_\odot$.  Besides, many ellipticals, presumably field
objects, are predicted to form at lower redshift, $z<2$.  This is
consistent with observations, which suggest that field ellipticals are
not a separate class of objects, but are younger on average (Bernardi
et al. 1998; Franceschini et al. 1998; Abraham et al. 1998; Manenteau
et al. 1998).  

The inversion of hierarchical order is in line with the general trend
of galaxy formation: the star-formation history at $z<1$ is dominated
by dwarf galaxies, while $L_*$ galaxies, both elliptical and spiral,
seem already in place at $z=1$ (see, e.g., Ellis 1998).  However, a
direct determination of a $z_f - M_H$ relation from the age of stars
in ellipticals is hampered by the well-known age-metallicity
degeneracy.  When trying to break this degeneracy, a possible age
dependence, consistent with the one proposed here is claimed by many
authors (Matteucci 1994; Bressan, Chiosi \& Tantalo 1996; Franceschini
et al. 1998; Caldwell \& Rose 1998; Ferreras, Charlot \& Silk 1998;
Pahre, Djorgovski \& de Carvalho 1998).

\subsection{Predictions on galaxies}

Given the number of galactic halos $n_H(M_H,p,z_{\rm sh})$
(Eq~\ref{eq:joint_all}, \ref{eq:ps}, \ref{eq:cutoff},
\ref{eq:spinpdf}, \ref{eq:merging}, \ref{eq:perc} and \ref{eq:dens}),
calculated at the shining time $z_{\rm sh}$, the total number of
galaxies at redshift $z=0$ with morphological type defined by $p_1\le
p \le p_2$ (the limits can depend on $M_H$) is readily calculated by
integrating the number density of halos in $p$ and $z_{\rm sh}$:

\bea \lefteqn{n_{\rm gal}(M_H)\; dM_H =}\label{eq:galaxies} 
\\&& \left(\int_0^\infty\, dz_{\rm sh}
\int_{p_1}^{p_2}\, dp\; n_H(M_H,p,z_{\rm sh}) \right) dM_H.  \nonumber
\eea

It is supposed that the halo spin or the merging fraction are
responsible for galaxy morphology.  Both mechanisms are likely to have
a role in determining whether a galaxy is going to be bulge-dominated:
galaxy mergers give rise to ``hot'' galaxies, and the profiles of big
ellipticals are consistent with the merging origin (see, e.g., Faber
et al. 1997).  On the other hand, low-spin systems can lead to large
bulge-disc ratios, as suggested, e.g., by Mo, Mao \& White
(1997)\footnote{Galaxy morphology has probably a more complex origin:
for instance, many lenticulars could come from spirals infalling into
cluster; see, e.g., Ellis 1998 and references therein.}.

Elliptical galaxies are assumed to be hosted either in low-spin
($\lambda\le \lambda_E$) halos or in halos which have suffered a major
merger ($1/2\le f\le f_E$).  In both cases, the probability for a halo
to host an elliptical increases with $M_H/M_*$.  As a consequence of
the inversion of hierarchical order for galaxy formation, described in
the previous subsection, small galactic halos shine later, when they
are small with respect to $M_*$ at their shining time.  Then, the
fraction of elliptical galaxies is not fixed but increases with mass;
this leads to a flattening of the elliptical mass function with
respect to the mass function of all the other halos.

With a fixed $p$-threshold and a suitable tuning of the free
parameters involved (see Section 2.7 for full details), it is possible
to obtain a satisfactory prediction for the halo mass function of
ellipticals, which is the main concern of the present paper.  However, in
Section 3.1 we will also test the predictions for the mass function of
spiral halos, so as to give further support to the galaxy formation
model presented here.

The mass function of non-elliptical halos is steeper than that of
spirals.  This is in line with the general behaviour of galaxy
formation models, which tend to predict a steep luminosity function
(see, e.g., White 1993; Cole et al. 1994).  This problem is usually
solved by assuming a low efficiency of star formation for small halos.
If spin is supposed to determine the galaxy type, it is possible to
subtract small-mass, high-spin halos by assuming that they are not
going to host bright galaxies but large low surface brightness (LSB)
discs.  Following Dalcanton, Spergels \& Summers (1997) and Jimenez et
al. (1997)\footnote{According to Jimenez et al. (1997), high-spin
halos do not even host LSB's, but remain dark.}, the spin threshold
$\lambda_S$ for the formation of non-bright galaxies is assumed to
depend on mass: small-mass halos are more likely to host a non-bright
galaxy.  The spin threshold is parameterized as follows:

\be \log\lambda_S (M_H) = \alpha_S \log(M_H/10^{12}\ M_\odot) + 
\log\lambda^0_S. \label{eq:lamlsb} \ee

\noindent
Then, a spiral halo is selected if $\lambda_E<\lambda\le
\lambda_S(M_H)$.  Consistency with observations is obtained for
$\alpha_S=0.4$ (see Table 2), in rough agreement with Jimenez et
al. (1997) who find $\lambda_S \propto M_H^{0.4}$, and Dalcanton et
al.  (1997) who report $\lambda_S \propto M_H^{1/6}$.

The predicted redshift evolution of the formation of bulge-dominated
galaxies can be tested through the star formation history of
elliptical galaxies given by Franceschini et al. (1998).  It is
possible to get a rough prediction of the contribution of ellipticals
to the star formation rate of the universe by assuming that they form
all their stars at the shining time of the halo, as defined in Section
2.4.  Assuming that the visible mass of an elliptical, $M_{\rm bul}$,
is connected to its light through the relation $M_{\rm
bul}/L=(M/L)_0(L/L_{*E})^ {\beta_{\rm bul}-1}$ (where $L_{*E}$ is the
Schechter parameter for the E luminosity function), the bulge mass is
related to the halo mass as:

\be M_{\rm bul} = \left(\frac{M}{L}\right)_0 L_{*E} \left(\frac{M_H}
{M_{H*}^E} \right)^{\beta_{\rm bul}/\beta_E}. \label{eq:bulge} \ee

\noindent
The parameter $\beta_E$ is the exponent of the $M_H-L$ relation for
ellipticals, and is defined in Section 3.1, Eq.~\ref{eq:msul}; its
assumed value is 0.75 (see also Table 1).  Following paper I, the
parameter $\beta_{\rm bul}$ is set to 1.25, while $(M/L)_0$ is set to
6.9$h$ (luminosities are in the $B$ band).  The star-formation rate is
then calculated as:

\bea \lefteqn{{\rm SFR}(z)\; dt=}\label{eq:sfr}\\&& 
\left(\int_{p\rightarrow E} dp  
\int_0^\infty dM_H M_{\rm bul}(M_H)\;  n_H(M_H,p,z) \right) 
\left|\frac{dz}{dt}\right| dt. \nonumber  \eea

\noindent
Here the integral in $p$ is performed over the interval which defines
the elliptical morphology.  

Eq.~\ref{eq:sfr} is only valid under the rather artificial hypothesis
that all stars in an elliptical form in a very short time.  This is a
good assumption only at low redshift, when the age of the Universe is
much larger than the typical duration of the starburst.  A better
estimate of the star-formation rate of ellipticals can be obtained by
convolving Eq.~\ref{eq:sfr} with a curve describing an average
star-formation history, given for instance by a truncated exponential
with timescale $\tau$:

\be {\rm SFR}'(z) dt = \int_{t+\tau}^\infty {\rm SFR}(t_{\rm sh})
\frac{1}{\tau}\exp\left(-\frac{t-t_{\rm sh}+\tau}{\tau}\right)dt_{\rm sh}
\label{eq:sfrp} \ee

\noindent 
In this case, the shining time is identified with the first $e$-fold
time, as the stabilization of the halo is supposed to be a result of
the massive star formation, and then cannot precede it.  In the
following we will use for $\tau$ a value of 1 Gyr, which is of order
of the delay time of a $10^{12}\ M_\odot$ halo.  Note that this is
probably larger than the star-formation timescale for giant
ellipticals, which is likely to be smaller than 0.3 Gyr (Matteucci
1994).  Of course, the star-formation timescale is physically related
to the delay time $t_{\rm delay}$, and is likely to depend on halo
mass.  However, an accurate modeling of such a timescale is beyond the
scope of the present paper.  Then, the convolved star-formation history
should be considered just as an indication of what can happen when
relaxing the hypothesis of very fast burst.  It is also noteworthy
that, as expected, the convolution does not change appreciably the
low-redshift cutoff of the star-formation history.

\subsection{Predictions on QSOs}

As mentioned in the Introduction, BHs strongly prefer elliptical
morphologies, as their masses correlate with the mass of the host
bulge.  This correlation presents a significant scatter of 0.3--0.5 in
decimal logarithm (see paper I).  This scatter reveals the need of a
``hidden variable'', able to modulate the efficiency of BH formation
in halos, thus giving the wanted broadness in the BH-bulge relation.
On the other hand, the mechanism responsible for galaxy morphology has
an indirect influence on BH formation.  As a reasonable working
hypothesis, we assume that both efficiency of BH formation and
galactic morphology are influenced by the same physical variable, spin
or merging.

The PDF of the morphological parameter must be such to reproduce,
under a reasonable transformation, the high luminosity tail of the QSO
luminosity function.  Although merging is a physically motivated cause
for stimulating the accretion onto a BH (it creates non-axisymmetric
perturbations which help the gas to fall toward the center of the
potential well), from the statistical point of view the merging
fraction $f$ results unsuitable to this purpose: as shown in Fig. 1,
the $f$-distribution is very flat around the value 0.5, which
corresponds to the merging of clumps of nearly equal mass, an event
which is not asymptotically rare.  Then, to shape this function into a
steep power-law, so as to fit the QSO luminosity function, it is
necessary to assume an extremely steep dependence of the efficiency of
BH formation on $f$.  This is unphysical, as the fate of gas is not
expected to be sensitive to infinitesimal variation of the ratio of
merging clumps.  However, the merging fraction is not the only
important physical quantity involved in a merging process.  Other
quantities, like the relative orientation of the spins of the merging
clumps, or their impact parameter, are likely to be important in
determining the final galactic morphology.  Then, a more detailed
modeling of the merging process is required to make it suitable for
modulating BH formation, but this is beyond the scope of the present
paper.

The spin of the halo does not suffer from the problem discussed above,
as low-spin halos are asymptotically rare.  Halo spin has already been
proposed as an important variable by Eisenstein \& Loeb (1995b).  On
the other hand, what is physically relevant for the formation of a BH
is the quantity of angular momentum that the gas is able to lose (see
also Cavaliere \& Vittorini 1998), and the internal distribution of
this angular momentum in the very center of the halo.  Moreover, with
a spin profile compatible with actual elliptical galaxies, it is very
difficult to have the formation of a huge BH (De Felice, Yu \& Zhou
1992).  Then, the relevance of the global spin of the DM halo for BH
formation is not obvious.  But the BH-bulge relation shows that the
final mass of the BH is related to some global property of the halo;
then, it is not unreasonable to assume it to be related to another
global property, as the total spin.  The physical cause of this
relation could be a dynamical feedback of the BH to the halo: a strong
central mass influences through chaotic mixing the orbits which pass
near the center, making them more axisymmetric (Merritt 1998).  This
could limit the mass of the BH, making it to depend on the total spin,
which is relevant for axisymmetric systems.  Alternatively, the
BH-bulge relation could be due to the competition with star formation
(Wang \& Biermann 1998) or to the mechanical feedback of the BH on the
protogalaxy (Silk \& Rees 1998); the latter mechanism can give a steep
dependence of the BH mass both on the bulge mass ($M_\bullet \propto
M_{\rm bul}^{5/3}$) and on the halo spin ($M_\bullet \propto
\lambda^{-5}$) (Haehnelt, Natarajan \& Rees 1998).  In any case, a
dependence of the BH mass on the total spin is expected if this
quantity influences the profile of the halo, and then its central
density.  Finally, as the halo spin is acquired from tidal torques
given by the large-scale structure, a direct influence of it on the BH
formation would explain the observational evidence of alignment
between radio loud AGNs, host elliptical galaxies and large-scale
structure (West 1994).

The mass of the BH changes rapidly in time during the bright QSO
phase.  However, as shown in paper I, the most important period, the
longest and the brightest, is that in which the BH has acquired most
of its final mass.  It is then reasonable to consider only the final
mass of the BH, $M_\bullet$, which is the result of the accretion onto
the seed BH of the matter available in the reservoir.  In the
following, the instant at which the BH acquires its final mass will be
called BH formation.  Further accretion of mass in a non-bright or
reactivated phase is assumed negligible.

To construct a prediction for the QSO luminosity function it is
necessary to specify the efficiency of BH formation in DM halos:

\be \varepsilon_H \equiv M_\bullet/M_H, \label{eq:eff} \ee

\noindent
i.e. the amount of matter which ends up into the BH in units of the
halo mass.  This is assumed to depend on spin in the following way:

\be \varepsilon_H = \varepsilon_{H0} \left(\frac{\lambda}{\lambda_0}
\right)^{-\alpha_q}. \label{eq:effspin} \ee

\noindent
The reference value $\lambda_0$ is set so as to be a 2$\sigma$ event
when $M_H=M_*$, while $\varepsilon_{H0}$ and $\alpha_q$ are left as
free parameters.  Eq.~\ref{eq:effspin} implies a linear scaling of
$M_\bullet$ with the halo mass $M_H$ (neglecting the weak mass
dependence induced by the spin).  However, BH masses appear to scale
with the bulge mass; this issue will be addressed in Section 5.3.

The number of BHs formed at redshift $z$ is:

\bea \lefteqn{n_\bullet(M_\bullet,z)dM_\bullet\; dz =} \label{eq:nbh} \\
&& \left(\int_0^\infty dM_H\; n_H(M_H,\lambda(M_\bullet,M_H),z) \right) 
\frac{dM_\bullet\; dz} {\alpha_q \ln 10}. \nonumber\eea

To obtain a QSO luminosity from a BH mass, it is assumed that the BHs
accrete mass at a fixed ratio of the Eddington rate, $f_{\rm ED} \equiv
L/L_{\rm ED}$, where the Eddington luminosity is defined as $L_{\rm ED}$ =
$4\pi G m_p c M_\bullet/\sigma_t$ $\sim$ $3.4\ 10^{4}
(M_\bullet/M_\odot)\ L_\odot$ ($m_p$ is the proton mass, $\sigma_T$
the Thompson cross section).  Then, the BH mass grows to its final
value during a timescale, or duty cycle time:

\be t_{\rm duty} = \varepsilon\, t_{\rm ED}/f_{\rm ED}, \label{eq:duty} \ee

\noindent
where the Eddington time is defined as $t_E= M_\bullet c^2/L_{\rm ED}$
$\sim 4\ 10^8$ yr and $\varepsilon$ is the efficiency of radiation of
the QSO in units of $Mc^2$, where $M$ is the accreted mass.  The
luminosity of the QSO in a given e.m. band is related to the BH mass
through:

\be L_{\rm QSO}(M_\bullet) =f_{\rm ED} L_{\rm ED}(M_\bullet) /C_B. \label{eq:lqso} \ee

\noindent
Here $C_B$ is the bolometric correction appropriate for the e.m. band
used.  Observational evidence (see, e.g., Padovani 1989; Wandel 1998)
suggests that the efficiency of accretion is a function of the QSO
luminosity, going from $\sim 0.05 - 0.1$ for small AGNs to $\sim 1$
for bright QSOs.  Following paper I, it is assumed that:

\be f_{\rm ED} = \left(\frac{L_{\rm bol}}{10^{49} {\rm
erg/s}}\right)^{\alpha_{\rm ED}},
\label{eq:edeff} \ee

\noindent
with the exponent $\alpha_{\rm ED}$ set to 0.2.  

The luminosity function of QSOs is then:

\bea \lefteqn{n_{\rm QSO}(L_{\rm QSO};z) dL =}\label{eq:qso}\\&&
n_\bullet(M_\bullet(L_{\rm QSO}),z)
\frac{M_\bullet}{L_{\rm QSO}}(1-\alpha_{\rm ED})\,
t_{\rm duty}\left|\frac{dz}{dt}\right| \; dL_{\rm QSO}. \nonumber \eea

\noindent It is noteworty that with the inversion of the hierarchical
order for galaxy formation, the ``inverted'' evolution of QSOs is
naturally obtained.

The predicted mass function of dormant BHs at $z=0$ is obtained simply
by integrating in redshift the number density of BHs given in
Eq.~\ref{eq:nbh}.

The bivariate number density of BHs hosted in bulges of a given mass
is obtained from Eq.~\ref{eq:joint_all}, by transforming the halo mass
and the spin into bulge (Eq.~\ref{eq:bulge}) and BH
(Eq.~\ref{eq:effspin}) masses, and by integrating the resulting
distribution in redshift:

\bea \lefteqn{n_{\bullet-{\rm bul}}(M_\bullet,M_{\rm bul})dM_\bullet\; 
dM_{\rm bul} =} \label{eq:corr} \\&& \left(\int_0^\infty dz\; 
n_H(M_H(M_{\rm bul}), \lambda(M_\bullet,M_H),z) \right) \nonumber \\&&
\frac{\beta_E}{\beta_{\rm bul} \alpha_q \ln 10} 
\frac{M_H(M_{\rm bul})}{M_{\rm bul}M_\bullet}
dM_\bullet M_{\rm bul}. \nonumber\eea

\noindent This equation is valid only for halos which host elliptical
galaxies.

Finally, while it is reasonable to assume that the global properties
of the galactic DM halo determine the amount of mass available to the
BH for accretion, the actual amount of mass accreted could depend on
some unpredictable details of the accretion process.  This could
influence the value of $f_{\rm ED}$, which would then become a random
variable (Siemiginowska and Elvis 1997; however, this would influence
the conclusions of paper I), or it could only influence the value of
$\varepsilon_{H0}$, which would then be modulated by a completely
random process.  These possibilities will be addressed elsewhere.

\subsection{Assumptions and parameters}

It is useful at this stage to list all the assumptions made in this
Section:

\begin{itemize}
\item The number of dark-matter halos dynamically formed at a given
redshift, with a given mass and a given spin or merging fraction, is
given by a set of ``numerical recipes'', mainly based on the extended
PS formalism, which are known to fit N-body simulations.
\item Galactic halos are distinguished from those corresponding to
galaxy groups and clusters according to a cooling criterion (which
suppresses galaxy formation in large halos) and a dissipationless
merging criterion (which suppresses galaxy formation at lower
redshift).
\item Because of feedback, galaxy formation (or, in other words, the
shining of the halo) is delayed with respect to the dynamical
formation of the halo.  The delay is larger for smaller halos.
\item The merging of galactic halos which have already
shone is neglected.
\item Elliptical galaxies are hosted either in low-spin halos or
in halos which have experienced a major merger.
\item High-spin halos do not host a bright galaxy.
\item The efficiency of BH formation in galactic DM halos is modulated
by the same physical variable which is responsible for the galactic
morphology.
\item The QSO activity is close in time to the main burst of star
formation for bulge stars.
\item QSOs shine only once, and acquire most of their mass in their
bright phase.
\item The efficiency of accretion of the BHs, expressed in Eddington
units, is a function of the BH mass; this is given in paper I.
\end{itemize}

The free parameters of the model are listed in Table 2, together with
the cosmological parameters which define the background cosmologies
used.  In the following Sections the parameters will be constrained by
comparing the predictions of the model to many distributions inferred
from observations: the mass function of galactic halos at low
redshift, divided in broad morphological classes, the star-formation
history of elliptical galaxies, the luminosity function of (optical
and obscured) QSOs and its evolution in redshift, the mass function of
dormant BHs in nearby galaxies, and the scatter in the BH-bulge
relation.  For each cosmology, it is possible to find a set of
acceptable parameters.

The best fit parameters are obtained by means of a qualitative
comparison of the predicted and observed quantities.  This is possible
because each parameter influences mainly some particular prediction, so
that they can be fixed separately.  In particular:

\begin{itemize}
\item the $\alpha_f$ parameter (Eq.~\ref{eq:delay}), which determines
the inversion of the hierarchical order for galaxy formation, is fixed
by reproducing the correct slopes of the galaxy mass functions;
\item the normalization time $t_f$ (Eq.~\ref{eq:delay}) is fixed by
reproducing the high-redshift evolution of the QSO luminosity
function;
\item the cutoff redshift $z_0$ (for Eq.~\ref{eq:dens}) is fixed by
reproducing the cutoff of the star-formation history of elliptical
galaxies;
\item $M_{\rm cool}$ (Eq.~\ref{eq:cutoff}) is fixed by fitting the cutoffs
of the galaxy mass functions;
\item $\lambda_E$ and $f_E$ (limits of the $p$-integral in
Eq.~\ref{eq:galaxies}) are fixed by reproducing the normalization for
the mass function of ellipticals;
\item $\lambda^0_S$ (Eq.~\ref{eq:lamlsb}) is fixed by reproducing the
normalization for the mass function of spirals;
\item $\alpha_S$ (Eq.~\ref{eq:lamlsb}) is fixed by reproducing the slope of
the spiral mass function;
\item the mass function of ellipticals in the merging case is
sensitive to the value of the timescale $\Delta \log b$
(Eq.~\ref{eq:merging}), which is set to 0.1, different values lead to
unsatisfactory mass functions;
\item $\alpha_q$ (Eq.~\ref{eq:effspin}) is tuned to obtain a good
shape for the QSO luminosity function;
\item $\varepsilon_{H0}$ (Eq.~\ref{eq:effspin}) is obtained by fitting
the mass function of dormant BHs;
\item $\varepsilon$ (Eq.~\ref{eq:duty}) is obtained by reproducing the
normalization of the QSO luminosity function once $\varepsilon_{H0}$
is fixed.
\end{itemize}

%%%%%%%%%%%%%%%%%%%%%%%%%%%%%  3  %%%%%%%%%%%%%%%%%%%%%%%%%%%%%%%%%%
\section{Galaxies}

\subsection{The halo mass function of galaxies}

Galaxies are embedded in DM halos, the mass of which is not directly
observable; the statistical quantity which can be observed is the
luminosity function.  However, at fixed morphological type galactic
halos are known to follow some kind of universal profile (Salucci \&
Persic 1998a).  This is theoretically confirmed by the existence of a
universal profile for DM halos (Navarro, Frenk \& White 1995; Moore et
al. 1998), even though observational and theoretical halos differ in
some details.  This fact implies that halo masses and galaxy
luminosities are strictly related through a morphology-dependent
$M_H/L$ relation, which is in general a function of luminosity.  Once
this relation is known, it is possible to determine the mass function
of objects from their luminosity function, and to compare them to the
predictions of the galaxy formation model of Section 2.

We divide the morphological types for bright galaxies into two broad
categories, one of early types (E and S0, briefly E), and one of
late types (Sa to I, briefly S).  The luminosity functions are
parametrized by means of the usual Schechter formula:

\be \phi_i(L)dL = \phi_{*i} (L/L_{*i})^{-\alpha_{i}}
\exp(-L/L_{*i})dL/L_{*i}, \label{eq:schechter}\ee

\noindent 
where the index $i$ is E or S.  Type-dependent luminosity functions
for E and S galaxies are given for instance by Efstathiou, Ellis \&
Peterson (1988), Loveday et al. (1992; Stromlo/APM survey), Marzke et
al.  (1994; CfA1+2 survey), Heyl et al.  (1997; the AUTOFIB survey),
Marzke et al. (1998; the SSRS2 survey), Marinoni et al. (1998, NOG
sample).  There is a broad agreement on the values of the various
parameters, but different authors disagree in some important details.
In particular, the E luminosity function may be flatter than the S
one, as suggested for instance by the early work of Efstathiou, Ellis
\& Peterson (1988), or by the AUTOFIB survey, but not confirmed by the
CfA1+2 or SSRS2 surveys.  A recent reinvestigation based on a large
local sample of galaxies (Marinoni et al. 1998) has confirmed that
the slope of the luminosity function steepens gradually from E to Sm/I
galaxies, with the exception of S0 galaxies, whose luminosity function
is as steep as that of Sc-Sd galaxies.  It is then apparent that the
true slope depends sensitively on the definition of the sample used.

In this work, values of 1 and 1.2 will be used for the slopes of the E
and S luminosity functions; these values are consistent with most
determinations.  Table 1 shows the values adopted for the parameters
of the E and S luminosity functions, which are roughly consistent with
all the luminosity functions listed above.

Following Salucci \& Persic (1998a), the mass-luminosity relations
are assumed to be of the kind:

\be M_H=M_{H*}^{i}(L/L_{*i})^{\beta_{i}}, \label{eq:msul}\ee

\noindent
where again $i$ is E or S.  Spiral galaxies have a $\beta_S$ parameter
of 0.56 and an $M_{H*}^S = 2.4\ 10^{12}\ M_\odot h^{-1} (\Omega(z_{\rm
sh})/\Omega_0)^{-1/3} (1+z_{\rm sh})^{-1}$ (Persic, Salucci \& Stel,
1996), where $z_{\rm sh}$ is the average shining redshif of the halo.
We take values 2.3, 2.5 and 3.0 for the EdS, Lambda and open models;
the results, reported in table 1, do not depend much on the exact
value of this parameter.  The $\beta_E$ and $M_{H*}^E$ parameters,
relative to early-type galaxies, are much harder to obtain, as the
evidence of DM is detected only in the outer regions (see, e.g.,
Danziger 1998).  The limited evidence available suggests a $\beta_E$
parameter not so different from $\beta_S$, and most likely smaller
than one (Salucci \& Persic 1998a).  The value 0.75 will then be used
in the following.  The $M_{H*}^E$ parameter is set equal to
$M_{H*}^S$.

\begin{table} \begin{center}
\begin{tabular}{l|c|c|c}
&  E   &  S\\ \hline \hline
$\phi_{*} (Mpc^{-3}h^3)$ & $3.8\ 10^{-3}$ & $1.1\ 10^{-2}$  \\
$\alpha $ & 1.0 & 1.2 \\
$M_B^* (mag-5\log h)$ & -19.8 & -19.8 \\
$M_{H*} (M_\odot)$, EdS, $h=0.5$& $1.45\ 10^{12}$ & $1.45\ 10^{12}$ \\
$M_{H*} (M_\odot)$, Lambda, $h=0.7$& $0.67\ 10^{12}$ & $0.67\ 10^{12}$ \\
$M_{H*} (M_\odot)$, open, $h=0.7$& $0.67\ 10^{12}$ & $0.67\ 10^{12}$ \\
$\beta$ & 0.75 & 0.56 \\ \hline
\end{tabular}\end{center}
\caption{Parameters for the galaxy luminosity functions and $M_H/L$
relations ($h=1$ unless otherwise stated).}
\end{table}

\begin{figure*}
\centerline{
\psfig{figure=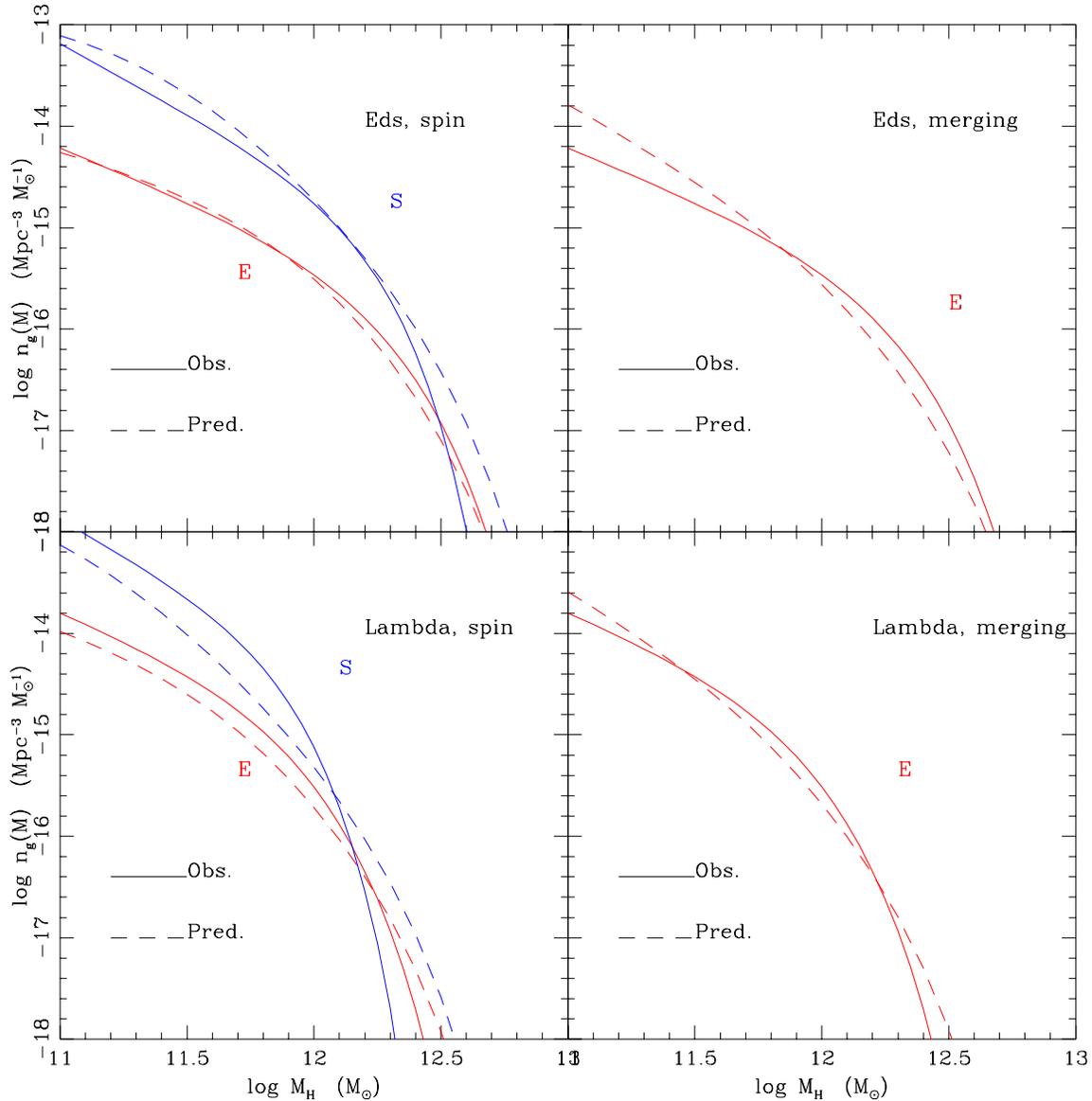,width=17cm}
}
\caption{Mass functions of galactic halos at $z=0$.}
\end{figure*}

\begin{table}
\begin{center}
\begin{tabular}{l|c|c|c|c|c|c}
& EdS & Lambda & open  \\ \hline \hline
Cosmological parameters \\ \hline
$h$               	   &  0.5 &  0.7  &  0.7  \\
$\Omega$          	   &  1   &  0.3  &  0.3  \\
$\Omega_\Lambda$  	   &  0   &  0.7  &  0  \\
$\Gamma$          	   &  0.5 &  0.21 &  0.21 \\
$\sigma_8$        	   &  0.7 &  1.0  &  1.0  \\ \hline

Model parameters: general \\ \hline
$\alpha_f$                 &  2.2 &  2.6  &  2.0  \\
$t_f (Gyr)$                &  0.2 &  0.5  &  0.6  \\
$z_0$                      &  1.0 &  0.7  &  0.7  \\
$\log M_{\rm cool}\ (M_\odot)$ & 12.3 & 12.1  & 12.1  \\ \hline

Model parameters: spin \\ \hline
$\log\lambda_E$            & -1.7 & -1.6  & -1.5  \\
$\log\lambda^0_S$          & -0.8 & -0.6  & -0.6  \\
$\alpha_S$         	   &  0.4 &  0.4  &  0.4  \\ 
$\alpha_q$                 &  1.8 &  1.3  &  1.7 \\
$\log\varepsilon_{H0}$     & -3.2 & -2.9  & -3   \\
$\log\varepsilon$          &  0.1 &  0.1  &  0.1 \\ \hline

Model parameters: merging \\ \hline
$\Delta\log b$             &  0.1 &  0.1 &  0.1 \\
$f_E$                      &  0.7 &  0.85 &  0.85 \\ \hline

\end{tabular}
\end{center}
\caption{Values of the free parameters for the cosmological models
considered. }
\end{table}

Fig. 3 shows the comparison of the predicted to the ``observed'' mass
functions in the cases of EdS and Lambda Universes (the open case is
very similar to the Lambda one, and is not shown).  Table 2 gives the
best-fit values of the parameters used, for the different cosmological
models.  Although ellipticals are the main concern of the present paper, we
show also the results for spirals in order to give further support to
our model and to our criterion for identifying the morphological type.

In the EdS case, the mass functions are reproduced with roughly the
correct slope and normalization.  The elliptical mass function
obtained with the merging fraction is steeper than the spin one.  The
exponential cutoffs are not perfectly reproduced, and in the spin case
ellipticals do not manage to outnumber spirals at large mass; this is
due to the fact that no inversion of hierarchical order is present at
large masses (Fig. 2).  We have chosen to tune $M_{\rm cool}$, so as
to best reproduce the mass function of ellipticals.

In the Lambda case, the overall normalization is slightly
underestimated.  However, the discrepancy is within the error in the
normalization of the observational mass functions, which is influenced
by the uncertainty in the model-dependent quantity $M_{H*}^E$, and by
the Hubble constant.  The predicted slopes for the mass functions
tend to be slightly steeper than in the EdS case, as the fraction
$M_H/M_*(z)$ varies more slowly in a non-critical Universe; this is
corrected by increasing the parameter $\alpha_f$ to 2.6.  Again, as
ellipticals are more relevant in this context, we have chosen to tune
the $\lambda_E$ and $f_c$ parameters so as to reproduce at best the
correct number of ellipticals, at the expenses of spirals.

The mass function of spiral halos (in the spin case) depends on the
subtraction of LSB halos, as explained in Section 2.5.  We have
verified that the predicted mass function for LSB is consistent with
the luminosity function given by Sprayberry et al. (1997) and the
$M/L$ ratio suggested by Salucci \& Persic (1998b).  These aspects of
galaxy formation will be discussed elsewhere.

It appears that both spin and merging give mass functions which are
roughly consistent with the ones inferred from observations.  This
implies that at this stage we do not need to choose between the two
mechanisms.  This conclusion will change in Section 4.

\subsection{The star-formation history of elliptical galaxies}

\begin{figure}
\centerline{
\psfig{figure=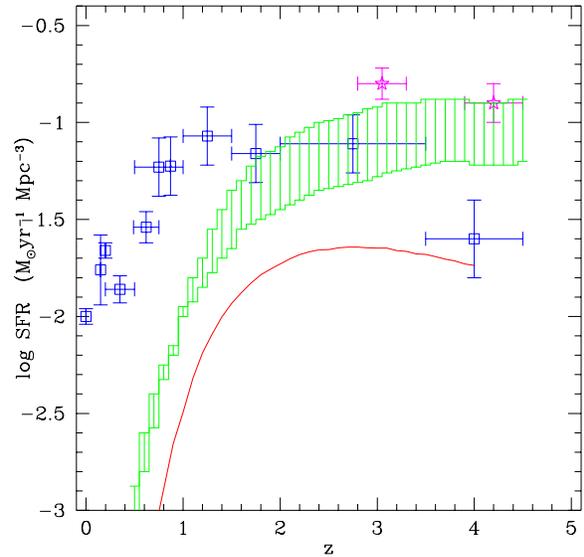,width=8cm}
}
\caption{Star formation history of elliptical galaxies, with an
assumed timescale $\tau$ of 1 Gyr (Eq.~\ref{eq:sfrp}), compared to the
result of Franceschini et al. (1998) (shaded region), and, for
reference, to the data of Madau (1997) (squares; corrected for dust
extinction as described in the reference) and Steidel et al. (1998)
(stars).}
\end{figure}

Fig. 4 shows for the EdS case the cosmological star formation rate of
ellipticals as inferred by the analytical model, assuming that bulge
stars form with a truncated exponential history with timescale 1 Gyr
(Eq.~\ref{eq:sfrp}; see Section 2.5).  It is worth recalling that this
curve is considered as just indicative of what happens when the
hypothesis of very fast burst is relaxed.  The prediction is compared
to the estimate of Franceschini et al. (1998), relative to ellipticals
in the Hubble deep field, and for reference to the estimates of the
global star-formation history by Madau (1997) (corrected for dust
extinction as discussed in the reference) and Steidel et al. (1998).

There is an inconsistency between the Franceschini et al. (1998) curve
and the bulge mass function given in paper I: the latter gives a mass
density of 6.3 10$^7\ M_\odot/Mpc^{-3}$ in bulge stars, while the
former gives from 1.2 10$^8$ to 2.5 10$^8$ in the same units.  The
model reproduces a halo mass function which is consistent with the
bulge mass function given in paper I, and then underestimates the
Franceschini et al.  curve by a factor from 2 to 4.  Taking this
inconsistency into account, the predicted star-formation history is in
agreement with that of Franceschini et al. (1998): it shows a decline
at $z<2$ and a very broad peak at $z\sim 3$, in rough agreement with
the flatness of the star-formation history suggested by Franceschini
et al. (1998) and Steidel et al. (1998) (see also Pascarelle, Lanzetta
\& Fernandez-Soto 1998), even though the normalization is not
recovered.  The star formation history of ellipticals will be
addressed in more detail in a forthcoming paper (Silva et al. 1999).

On the other hand, the star-formation history of all stars at $0<z<1$
is not reproduced; star formation is dominated at low redshift by
stars in spiral discs and in dwarf galaxies, i.e. objects which are
not related to the QSO phenomenon.  Note that this conclusion is at
variance with Boyle \& Terlevich (1998).

\subsection{Relationship with $z\sim 3$ galaxies}

In the last years, an increasingly large sample of galaxies at $z\sim
3$ has been observed; such galaxies are actively star-forming
galaxies, found as ``UV dropouts'' in deep fields (see, e.g., Steidel
et al. 1998).  These are often referred to as Lyman break
galaxies (hereafter LBG).  The clustering properties of such galaxies
are consistent with a scenario in which each galaxy is associated with
a single halo of $\sim 10^{12} M_\odot$ (Adelberger et al. 1998);
direct dynamical estimates of halo masses seem to indicate smaller
values, but the results are still too uncertain to draw firm
conclusions (Pettini et al. 1998).  It is interesting to note that the
clustering of QSOs, which evolves slowly with redshift, is more
similar to that of LBGs than to that of general galaxies (La Franca,
Andreani \& Cristiani 1998; Magliocchetti et al. 1998).  The abundance
of LBGs is $6.4\ 10^{-3} h^3 Mpc^{-3}$ for the EdS case, and $1.7\
10^{-3} h^3 Mpc^{-3}$ for an open model (Steidel et al. 1998).  Such
galaxies have been interpreted as the ancestors of big ellipticals
(Governato et al. 1998).

According to Fig. 2, LBGs would correspond to galaxies forming in
halos of $\ga 10^{12} M_\odot$, in broad agreement with the estimate
given above.  As a substantial fraction of $L_*$ galaxies are forming
at that redshift, the predicted abundance of such halos is of the
order of $\Phi_*$ for ellipticals; if it is assumed that only
bulge-dominated galaxies are actively forming stars at those
redshifts, then the abundance of halos able to host LBGs is broadly
compatible with the values given above (see table 1).  This implies
that LBGs do not sample the whole population of DM halos, but only
those whose characteristics (spin or merging in this context) are such
to cause strong star formation, while a majority of halos will host
proto-spirals which are not visible at that redshift.

In the present context, as long as each LBG is associated
to a single halo, these objects are going to survive to the present
epoch as big ellipticals; further mergings should be negligible in
terms of mass and star formation.  Then, the scenario presented here
is not in complete agreement with Governato et al. (1998), Kauffmann,
Nusser \& Steinmetz (1997) and Baugh et al.  (1998), who predict that
Lyman break galaxies are just pieces of big ellipticals, which must
merge subsequently.  The disagreement is weakened if the one-to-one
correspondence of halos and galaxies is relaxed; in this case the
merging of LBGs contained in a single halo does not imply any halo
merging.  An interesting example is given by the observation of the
distant radio galaxy 1138--262 (Pentericci et al. 1997, 1998): while
optical observations reveal a large amount of substructure,
corresponding to many (about ten) Lyman-break objects, observations in
the (rest frame) NIR and Lyman-$\alpha$ emission show a much
more coherent structure, with a velocity dispersion of $\sim 300\
km/s$.  This is consistent with a young giant elliptical, with some
knots of star formation visible in the optical as separated entities.
Multi-band observations of many elliptical galaxies are needed to
assess the dynamical state of high-redshift galaxies.

%%%%%%%%%%%%%%%%%%%%%%%%%%%%%  4  %%%%%%%%%%%%%%%%%%%%%%%%%%%%%%%%%%
\section{QSOs}

\subsection{Observational properties}

The optical luminosity function of QSOs is well known (see, e.g.,
Boyle, Shanks \& Peterson 1988).  The typical luminosity of QSOs
evolves rapidly with redshift: at smaller redshift, $z\la 3$, the
evolution is similar to a pure luminosity evolution with $L_*(z)
\propto (1+z)^{3.2}$ for $z<2$, and $L_*(z)\simeq {\rm const}$ at
$2<z<3$, while at $z>3$ QSOs start to decrease in number (see, e.g.,
Osmer 1998, Shaver et al. 1998).  At smaller redshift, $z\leq 1$, the
evolution is not purely in luminosity, as the luminosity function
tends to flatten (La Franca \& Cristiani 1997).

To describe the shape and evolution of the QSO luminosity function,
the parameterization proposed by Pei (1995) has been used (for a
spectral index $\alpha=-0.5$).  Luminosities are given in the $B$
band; following paper I and Elvis et al. (1994), the bolometric
correction has been set to 13.

We assume the existence of a whole population of obscured AGNs, which
contribute to the cosmological background in the hard X-rays (Setti \&
Woltjer 1989; Celotti et al. 1995; Comastri et al. 1995).  Following
Comastri et al. (1995), the abundance of such objects is estimated
starting from the soft-X-ray luminosity functions of QSOs (Boyle et
al. 1993).  It is assumed that for each observed soft-X-ray QSOs there
are 5.4 ones which are heavily obscured.  The applied bolometric
correction is 25 (see again paper I and Elvis et al. 1994).  It
results that obscured QSOs dominate the mass function of dormant BHs
at small masses, $M_\bullet<10^{8}\ M_\odot$, but do not contribute at
larger masses.

Figs. 5 and 6 show (for the EdS and Lambda models) the optical
luminosity function, the estimated contribution of obscured objects
and their sum; both are given in terms of the bolometric luminosity.
Although the calculations have been performed using the analytical
parametrizations, the data points from Pei (1995) and Boyle et
al. (1993) are shown in the figure.  At $z=4.4$, the newer data points
of Kennefick, Djorgovski and Meylan (1996) are reported in place of
the Pei data at $z=4$.  Obscured objects contribute significantly only
in the low-luminosity end, while bright QSOs are almost unobscured.
The contribution of obscured QSOs is not considered at $z>3$, as the
X-ray luminosity function is not measured there.  It is interesting to
see that the composite luminosity function appears almost featureless
and quite steep at all luminosities.  In this case, the knee of the
optical luminosity function would be interpreted as an effect of the
onset of obscuration.  It must be noted that both luminosity functions
are still uncertain, which makes the complete luminosity function
still speculative.  Nonetheless, the qualitative trends of the
dominance of obscured objects only at small luminosities and the
steepening of the complete luminosity function should be robust.

The evolution of the luminosity function is quantified, following Pei
(1995), through the index $\langle L_{\rm bol}^2 \rangle$ as a
function of redshift.  The exponent 2 is such to give more weight to
the objects near the knee of the luminosity function, which is most
robust; in this way the evolution index is not sensitive to errors at
both ends of the luminosity function.  This index is not sensitive to
the contribution of obscured objects, which is then neglected.  The
evolution index $\langle L_{\rm bol}^2 \rangle$ is shown in Fig. 7.

\begin{figure*}
\centerline{
\psfig{figure=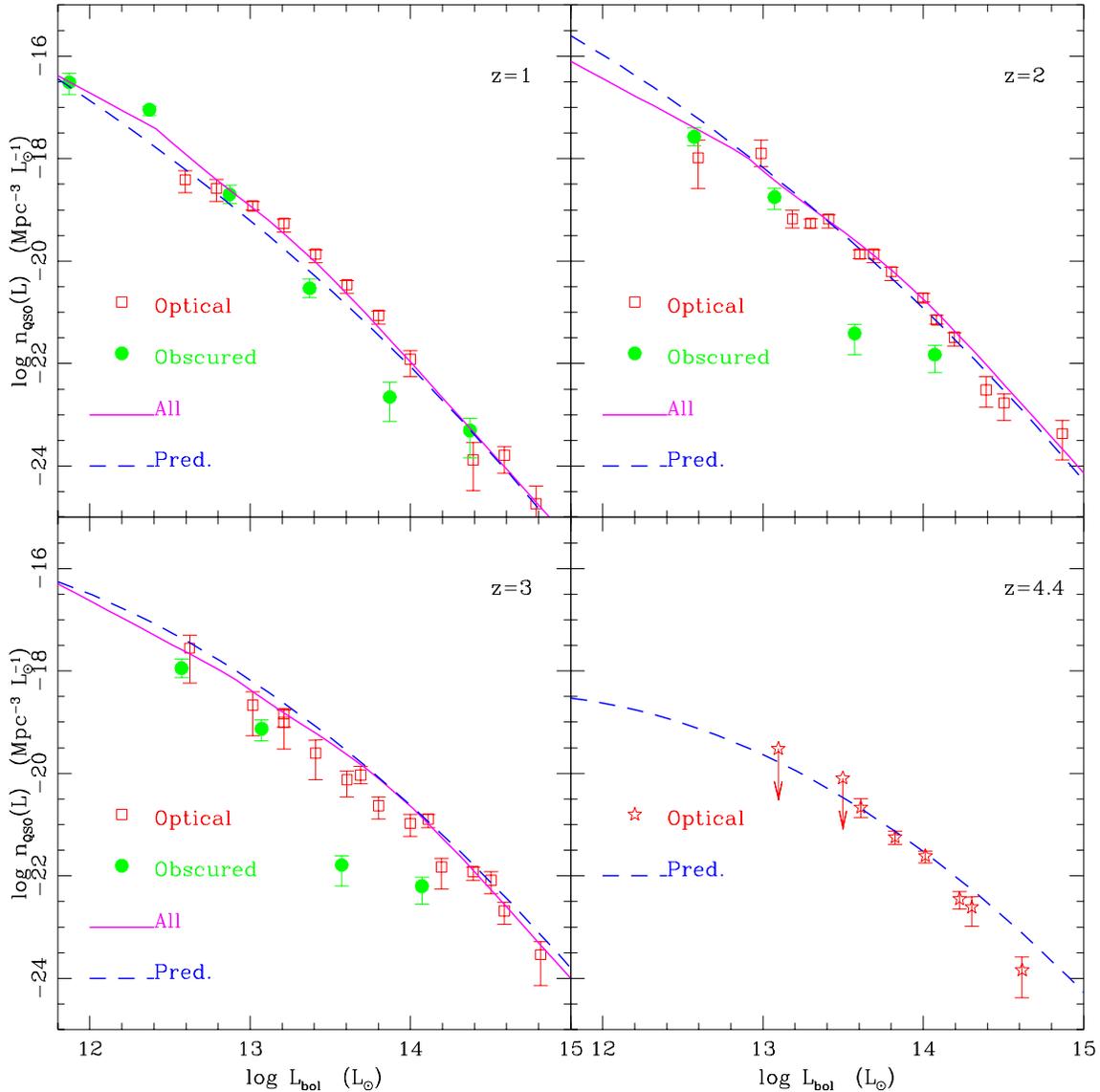,width=17cm}
}
\caption{QSO luminosity functions at different redshift. EdS case.
The optical luminosity function is taken from Pei (1995), the
contribution of obscured objects is based on the X-ray luminosity
function of Boyle et al. (1993).  The curve named ``ALL'' gives the
total contribution of optical and obscured objects, and is based on
the parametrizations of the optical and X-ray luminosity functions
given by the already mentioned authors. At $z=4.4$ the newer data
points of Kennefick et al.  (1996) (denoted by stars) are reported.}
\end{figure*}

\begin{figure*}
\centerline{
\psfig{figure=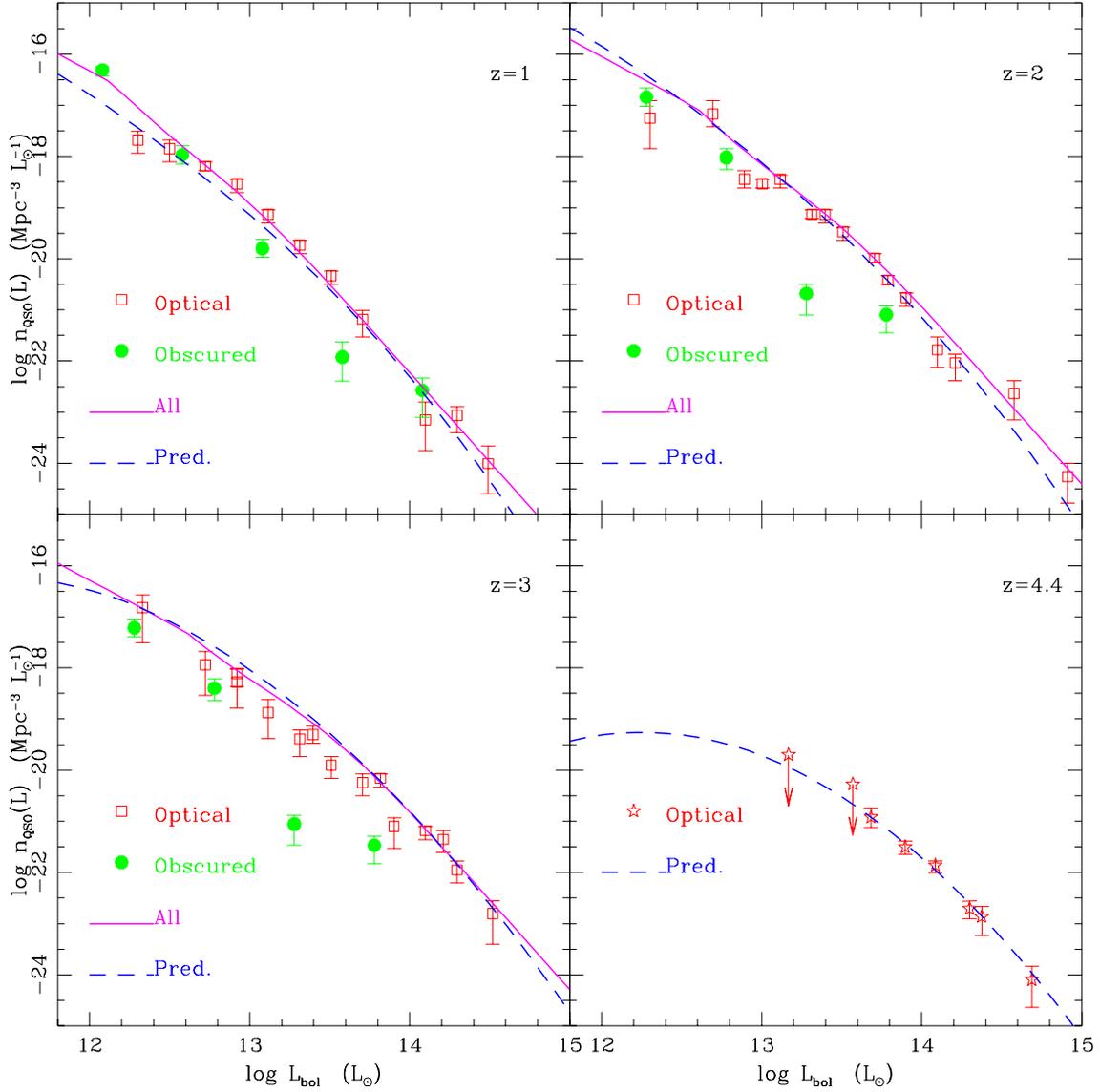,width=17cm}
}
\caption{The same as Fig. 5 for the Lambda case.}
\end{figure*}

\begin{figure*}
\centerline{ \psfig{figure=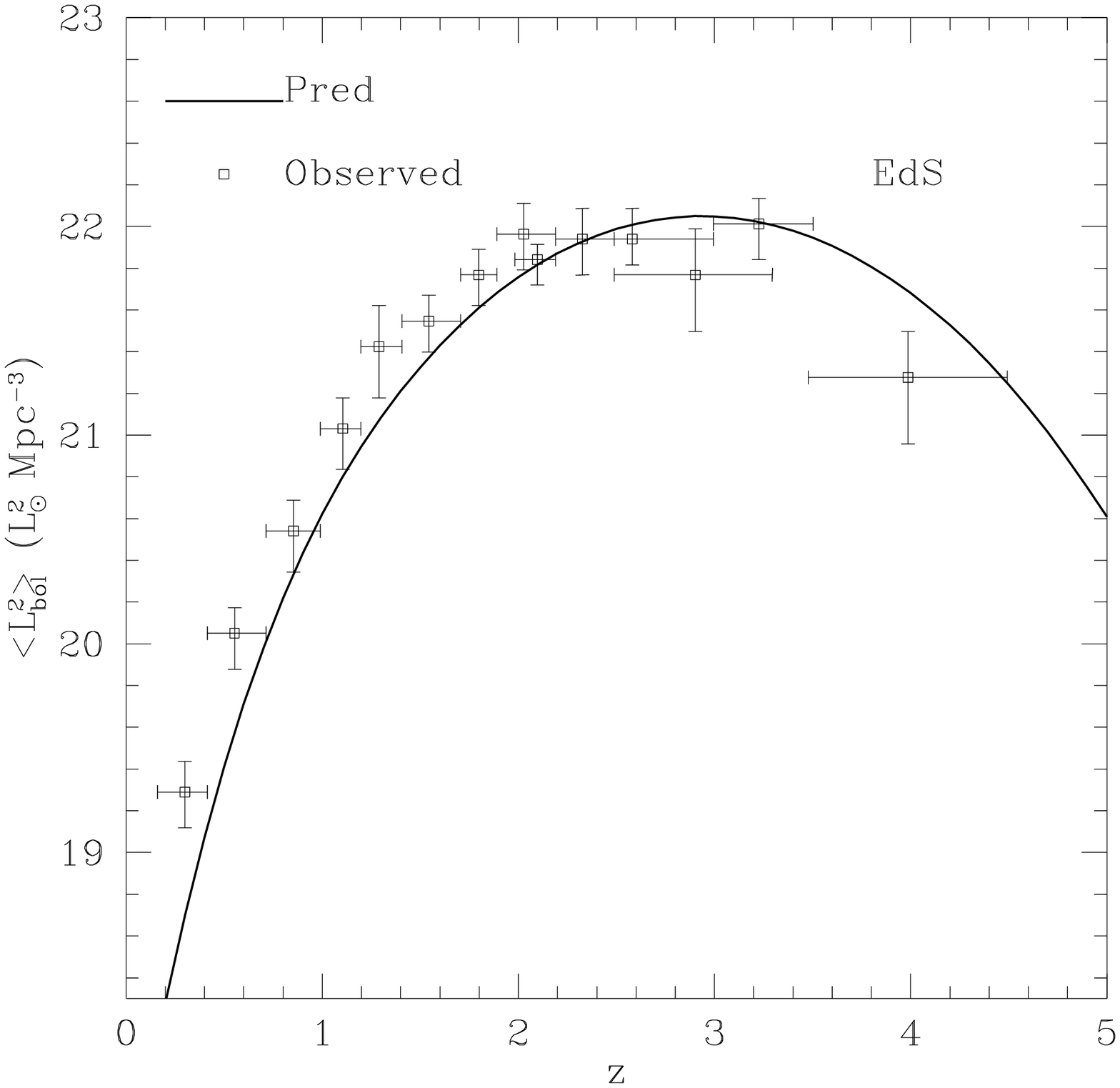,width=8cm}
\psfig{figure=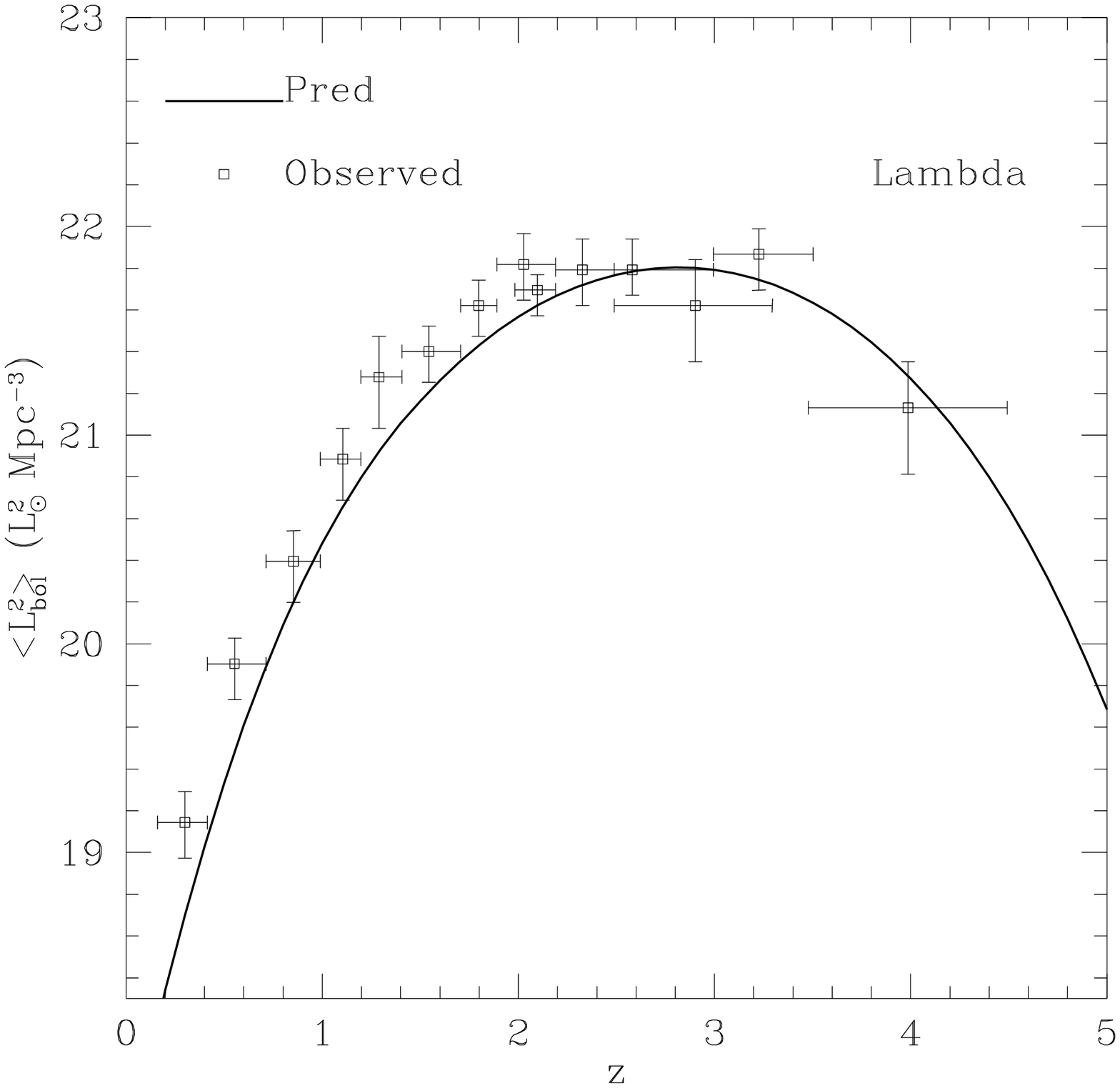,width=8cm} }
\caption{Evolution index $\langle L_{\rm bol}^2 \rangle$ for the QSO
population.  Data are taken from Pei (1995). Left: EdS case; right:
Lambda case.}
\end{figure*}

\subsection{Comparison of predictions with observations}

Figs. 5, 6 and 7 show the comparison of the predicted and observed QSO
luminosity functions and evolution index $\langle L_{\rm bol}^2
\rangle$, for the EdS and Lambda models.  The free parameters involved
are $\varepsilon$, $\alpha_q$ and $\varepsilon_{H0}$ (Section 2.6),
together with $t_f$ and $\alpha_f$ (Section 2.4); the best-fit values
are again given in Table 2.  The parameters have been chosen so as to
fit the complete luminosity function; as the predicted curve is almost
featureless (it is approximately a rescaling of the low-spin tail of
the lognormal spin PDF), it is much easier to fit satisfactorily the
complete luminosity function rather than the optical one only; anyway,
a moderately good fit of the optical curve can be obtained by
increasing the value of the $\alpha_q$ parameter.

The agreement between model and data is overall very good.  At smaller
redshift, $z< 1$, the number of QSOs is slightly underestimated,
especially at small luminosities.  This modest disagreement may imply
that recurrency of AGN activity is present at small redshift.  At high
redshift the model predicts a luminosity function which is flatter
than the one extrapolated from lower redshift, while lower-luminosity
activity is severely suppressed.  This is consistent with the
currently available upper limits of Kennefick et al. (1996), although
the EdS prediction tends to overpredict the high-luminosity tail.  The
open model, not shown for brevity, tends to predict a larger number of
high redshift QSOs, unless the delay time $t_f$ is more than 0.6 Gyr.

The parameter $\alpha_q$ is set to 1.8, 1.3 and 1.7 in the EdS, Lambda
and open cases, confirming that spin should play a major role in BH
formation, but it cannot be as large as 5, the value suggested by
Haehnelt et al. (1997).  The efficiency of radiation turns out to be
equal to the canonical value 0.1; this is a good consistency test for
the model.

%%%%%%%%%%%%%%%%%%%%%%%%%%%%%  5  %%%%%%%%%%%%%%%%%%%%%%%%%%%%%%%%%%
\section{Dormant black holes}

\subsection{The mass function of dormant BHs}

The determination of the mass function of dormant BHs in ordinary
galaxies at $z=0$ has been addressed in full detail in paper I, and
briefly described in the Introduction.  Here we present a comparison
of the mass functions as obtained from the QSO luminosity function and
the radio luminosity function of elliptical cores.  The mass function
of dormant BH masses is obtained from the QSO luminosity function (it
is called AMF, as in paper I) under the same assumptions described in
Section 2.6, i.e. of accretion at $f_{\rm ED}$ (Eq.~\ref{eq:edeff})
times the Eddington limit for a time $t_{\rm duty}$
(Eq.~\ref{eq:duty}):

\bea \lefteqn{n_\bullet(M_\bullet)dM_\bullet =\frac{\ln
10}{(1-\alpha_{\rm ED})} \frac{L}{t_{\rm duty}} }\label{eq:dormobs}\\&&
\left(\int_0^\infty dz \left| \frac{dz}{dt} \right|^{-1} n_{\rm QSO}
(L_{\rm QSO}(M_\bullet);z) \right) d\log M_\bullet.  \nonumber \eea

\noindent
Obscured QSOs have been included by using the complete luminosity
function for $n_{\rm QSO}(L;z)$.

\begin{figure*}
\centerline{
\psfig{figure=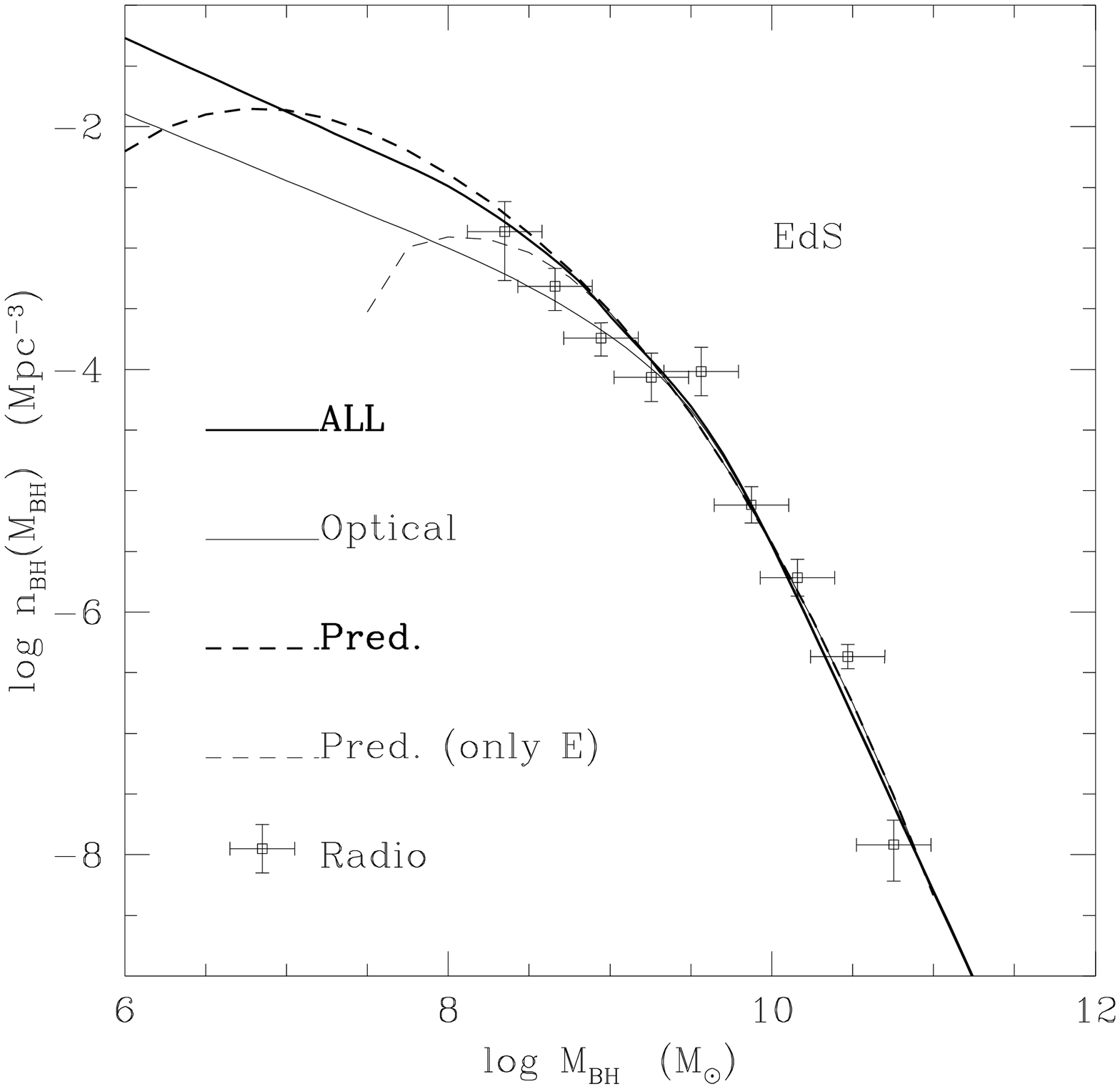,width=8.5cm}
\psfig{figure=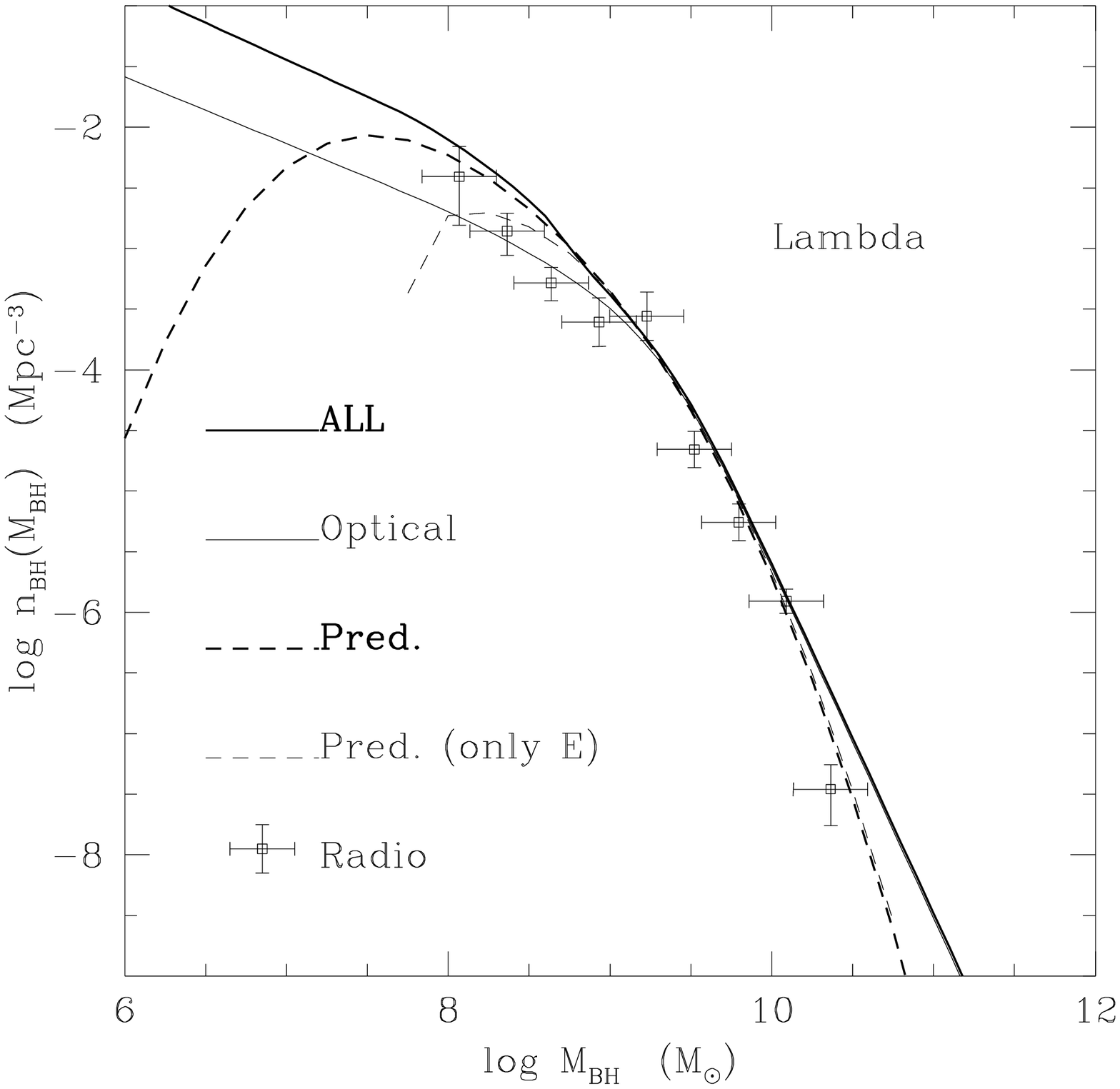,width=8.5cm}
}
\caption{Mass functions of dormant BHs.  The distribution
$n_\bullet(M_\bullet)$ is in $d\log M_\bullet$.}
\end{figure*}

The radio-based mass function (called RMF) was obtained, as in paper
I, by transforming the radio core luminosity function of ellipticals
(by Sadler et al. 1988, corrected to subtract non-core emission as in
paper I).  The mass function obtained in paper I (OMF in that paper)
from the bulge mass function is not shown, as it is trivially
satisfied when both the mass function of elliptical galaxies and the
efficiency of BH formation (Section 5.2) are correctly reproduced.

Fig. 8 shows the comparison between the predicted BH mass function and
the AMF and RMF.  Again, the EdS and Lambda models are shown.  The
agreement is again very good.  This result is trivial once the QSO
luminosity function is correctly reproduced at any redshift, and once
the consistency of the AMF and RMF has been assessed in paper I.
However, fitting the AMF and RMF allows one to determine the parameter
$\varepsilon_{H0}$ separately from $\varepsilon$, as this quantity
does not depend on the light actually emitted by the QSO.

Fig. 8 shows also the predicted contribution to the BH mass function
of elliptical galaxies alone (defined through the spin threshold),
which is the quantity to be compared to the RMF; the contributions
with and without ellipticals differ where the RMF is not defined.  As
in paper I (see their Fig. 5), ellipticals give the dominant
contribution to the mass function for $M_\bullet\ga 10^8\ M_\odot$,
while spirals dominate at smaller BH masses.  This gives further
support to the threshold criterion used to separate ellipticals from
spirals.

\subsection{The $M_\bullet/M_{\rm bul}$ ratio.}

A more interesting test would rely on the prediction of the joint
number density of BH and bulge masses, given by Eq.~\ref{eq:corr}.
This bivariate distribution gives full information on the correlation
between BH and bulge masses, and could be compared to direct estimates
of BH masses in nearby early-type galaxies.  However, the available
samples of galaxies with known BH masses are just compilations of
galaxies for which suitable observations were obtainable, and the
dynamical measures of BH masses are still affected by systematic
uncertainties connected to the kinematical models used (Magorrian et
al. 1998; van der Marel 1998; Ho 1998).  As a consequence, it is not
possible at present to obtain a reliable observational estimate of the
bivariate $M_\bullet-M_{\rm bul}$ distribution.

A more robust test relies on predicting the PDF of the quantity:

\be R \equiv M_\bullet/M_{\rm bul}. \label{eq:ratio} \ee

\noindent
The ratio $R$ is computed from the joint number density of BH and
bulge masses (Eq.~\ref{eq:corr}) by integrating it over one variable.
The PDF of $R$ has been estimated by Magorrian et al. (1998), on the
basis of their sample of BH masses; they fit their PDF with many
functions, among which a truncated decreasing power-law and a
lognormal turn out to be acceptable fits.  Paper I gives a different
estimate of this PDF: assuming a lognormal shape, the distribution
must be such to obtain a good BH mass function from the mass function
of galaxy bulges (the OMF), consistent with the AMF and RMF.  The
obtained values for the width and mean of the lognormal are 0.3 (in
decimal logarithm) and $10^{-2.6}$; the width is slightly smaller than
the one recovered by Magorrian et al. (1998) (0.5 in decimal
logarithm), implying that random errors in the estimates of BH masses
are not a main source of the scatter in the $M_\bullet - M_{\rm bul}$
relation.  As already mentioned in the Introduction, the average
obtained in paper I is smaller than the Magorrian et al.'s value by a
factor of $\sim 2$, in agreement with van der Marel (1998) and Ho
(1998), and this is commented in full detail in paper I.

The BH mass has been supposed to scale with the halo mass, which
implies a BH-bulge relation with slope $\beta_E/\beta_{\rm bul}\sim
0.6$.  As a consequence, the $R$ PDF depends much on the range of
bulge masses over which it is averaged.  The estimate of the $R$ PDF
given in paper I is sensitive only to the large-mass part of the mass
function, as it is designed to spread the sharp cutoff of the bulge
mass function into the milder cutoff of the BH mass function.
Besides, the BH-bulge correlation is best tested for big ellipticals,
while at lower bulge masses the measures are very uncertain and spiral
bulges come into play.  Then, the predicted $R$ PDF is calculated only
for ellipticals corresponding to luminosities larger than $L_*$.
Fig. 9 shows the comparison of the predicted $R$ PDF and the one
estimated in paper I.  In both the EdS and the Lambda cases, the peak
is reproduced at the correct position, while the distribution is
skewed toward large $R$ values and truncated at low $R$ values by the
spin threshold.  In case of poor and noisy data, this distribution
would be easily fit by the lognormal given in paper I.  Notably, at
large $R$ values the slope of the PDF is consistent with that of the
truncated power-law of Magorrian et al. (1998), which is also shown in
Fig. 9.  It must be stressed that at this stage the $R$ PDF is
reproduced without tuning any parameter.  The parameter which mostly
influences the shape of this curve is $\alpha_q$, which is already
constrained by the QSO luminosity function.  In other words, a
connection is estabilished between the shape of the QSO luminosity
function and the shape of the $R$ PDF.

\begin{figure}
\centerline{
\psfig{figure=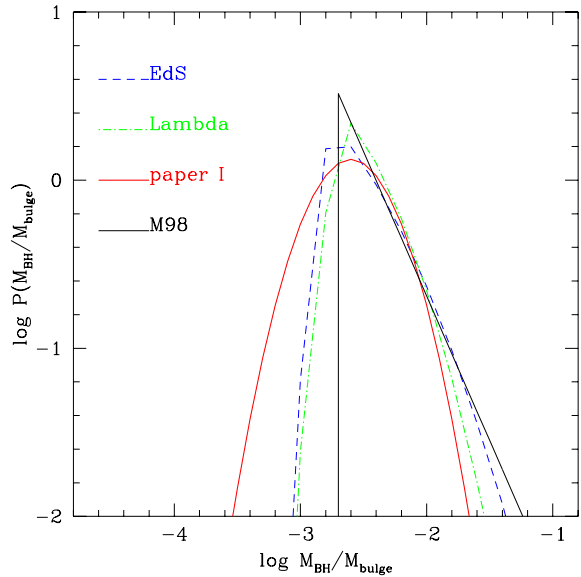,width=8cm}
}
\caption{PDFs of the $R$ ratio defined in Eq.~\ref{eq:ratio}.  The
thick continuous lognormal curve is obtained from paper I, while the
thin continuous line is the power-law fit of Magorrian et al. (1998).
The EdS and Lambda curves are shown.}
\end{figure}

As mentioned above, the mass of formed BHs has been supposed to scale
linearly with the halo mass, and this implies a BH--bulge relation
with slope $\sim 0.6$.  On the other hand, the data of Magorrian et
al. (1998) seem to suggest a linear or steeper relation between BH and
bulge mass, while van der Marel (1998) suggests that the relation
could be shallower than linear, in line with our prediction.  To force
a linear BH--bulge relation, one can simply assume that the efficiency
of BH formation scales with the bulge mass.  This is implemented by
multiplying the right-hand-side of Eq.~\ref{eq:effspin} by a term
$(M_H/(10^{12}\ M_\odot))^{\alpha_H -1}$, so that BH masses scale as
$M_H^{\alpha_H}$, then setting $\alpha_H=\beta_{\rm bul}/\beta_E$.
With this hypothesis, it is possible to reproduce satisfactorily the
QSO luminosity function and dormant BH mass function; in the EdS model
this is done by lowering the $\alpha_q$ parameter to 1 and the
$\varepsilon_{H0}$ parameter to $10^{-3.5}$.  However, the $\alpha_q$
parameter being lowered, the resulting $R$ PDF is significantly
narrower than those shown in Fig. 9, and then not compatible with the
observational evidence.

Then, in the present framework a linear scaling of $M_\bullet$ with
$M_{\rm bul}$ is not consistent with the data, unless a significant
part of the scatter in the $M_\bullet-M_{\rm bul}$ correlation is due
to observational errors, or unless a further mechanism is able to
increase the scatter in the BH-bulge relation, without changing the
QSO luminosity function; dissipationless galaxy mergers which take
place after the QSO phase could provide such a mechanism.  An even
steeper relation, $M_\bullet \propto M_{\rm bul}^{5/3}$, is suggested
by Silk \& Rees (1998); such a steep dependence would be hard to
reconcile with the framework presented here.

%%%%%%%%%%%%%%%%%%%%%%%%%%%%%  6  %%%%%%%%%%%%%%%%%%%%%%%%%%%%%%%%%%
\section{Summary and conclusions}

This paper describes an analytical model which addresses for the first
time the joint formation of galaxies and QSOs.  The model is able to
predict the halo mass function of galaxies of different broad
morphological types, the star-formation history of elliptical
galaxies, the QSO luminosity function and evolution, the mass function
of dormant BHs in nearby galaxies, and the correlation of these with
the mass of host bulges.  The model has been successfully compared to
available observational data, so as to constrain its free parameters
(listed in Table 2).  An acceptable fit is found for each of the
cosmological models considered (EdS, Lambda, open).

The following conclusions can be drawn:

\begin{itemize}
\item 
Consistency with observation is obtained if the hierarchical order is
inverted when considering the shining epoch of galactic halos: large
galactic halos experience massive star formation and QSO activity
before smaller ones. This is achieved by delaying the shining of small
objects.  The inversion of hierarchical order is consistent both with
the evolution of QSOs and with evidences based on stellar populations
of elliptical galaxies.
\item
There is a clear need of a ``new variable'', which modulates the
efficiency of BH formation for halos of a given mass.  As QSOs prefer
early-type morphologies, it is assumed that the ``new variable'' is
the same as the one responsible for galaxy morphology.
\item
The spin of DM halos is a good candidate as ``new variable''.  The
merging fraction appears a good candidate for determining galaxy
morphology, but not for modulating the efficiency of BH formation, as
the merging of halos of similar size is not asymptotically rare as
bright QSOs are.  A more detailed description of mergers could solve
this problem.
\item
The inversion of the hierarchical order of galaxy formation, together
with a threshold criterion for predicting morphological types (based
either on spin or merging) produces mass functions for galactic halos
which have the correct morphology-dependent slopes.
\item
A connection is estabilished between the slope of the QSO luminosity
function and the distribution of the ratio $M_\bullet/M_{\rm bul}$.
\end{itemize}

A {\it caveat} on the role of spin is necessary: as only the
statistical properties of spin have been used in the model, the
results presented here do not give direct evidence that spin is in
play in determining the luminosity of QSOs, but reveal that any
physical variable which has the same behaviour as the spin, i.e. has a
PDF nearly lognormal in shape slowly changing with $M_H/M_*$, is a
viable variable for both determining the morphological type and
modulating the mass of the BHs.

This paper gives further support to the favoured scenario of paper I,
in which the most luminous QSOs (associated to BHs with masses larger
than $10^8\ M_\odot$) are hosted in elliptical galaxies, shine only
once for an Eddington time and at the Eddington limit, and are hardly
obscured, while fainter AGNs are associated to spirals, shine at a
fraction of the Eddington limit, may be significantly reactivated and
are often heavily obscured.

The analytical model presented here is based on a number of reasonable
but simplified assumptions.  A more detailed description of DM halos
would require the use of semi-analytic merging trees, or of large
N-body simulations.  Merging of galactic halos after the QSO phase is
neglected by construction.  Spin and merging are treated as
alternative quantities, while they are likely to have both a role in
shaping galaxies and triggering BH formation.  The reactivation of
existing QSOs, important to reproduce the low-level activity at low
redshift, is neglected.  Nonetheless, the analytical model is supposed
to catch the most important elements in the process, and its agreement
with many different pieces of observational evidence is encouraging.

Our analysis shows that a delay of the ``shining phase'' of halos,
when star formation and QSO activity occur, makes it possible to
explain the main statistical features of the joint QSO-galaxy
formation.  This delay leads to an inversion of the hierarchical order
for galaxy formation.  This highlights the potential impact that QSOs
can have in galaxy formation.  Their importance relies on the ever
growing evidence that QSO activity is intimately related to galaxy
formation; the BH-bulge relation is a striking demonstration of such a
relationship.  The inversion of hierarchical order is not in
contradiction with standard hierarchical CDM models, as it is related
not to the dynamical formation but to the shining phase of DM halos,
and it is supposed to be caused by feedback mechanisms.  The inversion
of hierarchical order leads to the prediction that smaller galaxies
are made up of younger stars.  This prediction can be tested by
observations which are able to break the well-known age-metallicity
degeneracy which affects old stellar populations; the evidence already
available is consistent with it (Matteucci 1994; Bressan, Chiosi \&
Tantalo 1996; Franceschini et al. 1998; Caldwell \& Rose 1998;
Ferreras, Charlot \& Silk 1998; Pahre, Djorgovski \& de Carvalho
1998).

\section*{Acknowledgements}

The authors thank Ewa Szuszkiewicz for discussions.  P.M. thanks
Bianca Poggianti, Neil Trentham, George Efstathiou, Alfonso Cavaliere
and Martin Rees for discussions.  P.M. has been supported by the EC
Marie Curie TMR contract ERB FMB ICT961709.

\bsp

\label{lastpage}

\end{document}